\documentclass{jfm}
\usepackage{graphicx}
\usepackage{epstopdf,epsfig}
\usepackage{amssymb}
\usepackage{amsmath}
\usepackage{caption}
\usepackage{subcaption}
\usepackage{bm}
\usepackage{adjustbox}
\usepackage{amsmath}
\usepackage[percent]{overpic}

\usepackage{natbib}
\usepackage{hyperref}
\hypersetup{
    colorlinks = true,
    urlcolor   = blue,
    citecolor  = black,
}

\newcommand{\numtag}[1]{\tag{\theequation#1}}

\newcommand{\RomanNumeralCaps}[1]
\linenumbers
\graphicspath{{fig/eps/}}

\title{A framework for diagnosing inertial lift generation in wall-bounded flows: application to eccentric rotating cylinders in Newtonian and shear-thinning fluids}

\author{Masafumi Hayashi\aff{1}
  \corresp{\email{hayashi.masafumi.5g4@ecs.osaka-u.ac.jp}},
  Kazuyasu Sugiyama\aff{1}$^,$\aff{2}}

\affiliation{\aff{1}Graduate School of Engineering Science, The University of Osaka, Toyonaka, Osaka 560-8531, Japan
\aff{2}RIKEN, Wako, 2-1 Hirosawa, Saitama 351-0198, Japan}

\begin{document}
\maketitle

\begin{abstract}
\end{abstract}
A body moving in a wall-bounded flow often experiences a hydrodynamic lift force normal to the wall, which plays an important role in many fluid systems. 
In this study, we develop a framework for diagnosing steady inertial lift from the internal structure of the flow field. 
Based on the generalised reciprocal theorem for finite-Reynolds-number flows, the lift is expressed as a volume integral that identifies both the dominant contributions and the regions from which they arise. 
We apply this framework to numerically obtained steady flows of Newtonian and shear-thinning fluids between eccentric rotating cylinders, and analyse the lift acting on the inner cylinder undergoing rotation and orbital motion. 
In particular, we focus on lift reversal induced by increasing eccentricity in a Newtonian fluid and on lift reversal induced by stronger shear-thinning behaviour at high eccentricity.
The volume-integral expression decomposes the lift into a vortex-force contribution associated with inertia and a viscous stress contribution associated with the non-uniform viscosity field, and shows that the former dominates over the parameter range considered here. 
As the eccentricity increases, negative relative vorticity, and in some cases tangential velocity, become stronger in the narrow-gap region, thereby enhancing the negative local vortex-force contribution and inducing lift reversal.
Stronger shear-thinning behaviour, on the other hand, amplifies negative relative vorticity near the inner cylinder, thereby increasing the positive local vortex-force contribution and inducing lift reversal. 
These results demonstrate that the proposed framework is useful for diagnosing and interpreting steady inertial lift in wall-bounded flows.
\begin{keywords}
  Particle/fluid flow, Rotating flows, Rheology
\end{keywords}


\section{Introduction}
A body moving in a wall-bounded flow often experiences a hydrodynamic lift force in the wall-normal direction.
This lift plays a role in a wide range of fluid phenomena, such as the passive focusing of cells in microfluidic devices~\citep{DiCarlo2009Inertial_ed,Zhou2013Fundamentals_yu,Zhang2016Fundamentals_kw}, particle deposition in slurry flows in pipelines~\citep{alam2016slurry,zhang2016study,sinha2017review}, and lateral vibration of rotors in bearing systems of rotating machinery~\citep{newkirk1925shaft,benckert1980flow,muszynska1986whirl}.
Understanding the mechanism of lift generation in wall-bounded flows is therefore important for improving the performance of such fluid systems and for controlling their behaviour.

In the present study, we focus on steady inertial lift arising in wall-bounded flows at finite but small Reynolds numbers.
We also examine how rheological effects, particularly shear-thinning behaviour, modify this inertial lift.
In general, inertial lift in wall-bounded flows originates from symmetry breaking in the system.
In the Stokes limit, time-reversal symmetry (kinematic reversibility) holds, and the lift therefore vanishes in systems with fore-aft symmetry~\citep{bretherton1962motion}.
By contrast, at finite Reynolds numbers, time-reversal symmetry is broken by the nonlinearity associated with inertia, and a force in the wall-normal direction arises~\citep{ho1974inertial,vasseur1976lateral}.
For a better understanding of inertial lift in wall-bounded flows, 
it is important to clarify what kind of asymmetry forms in the flow field and how it is connected to the wall-normal force.
However, it is not straightforward to interpret physically how such lift is generated, especially at low Reynolds numbers.
Mechanically, lift is expressed as the integral of the fluid stress acting on the body surface, namely the pressure and shear stress, and it is thus natural to first examine the surface-stress distribution when investigating its origin.
However, especially at low Reynolds numbers, lift is often very small compared with drag~\citep{shi2020lift}, and the asymmetry appearing in the surface-stress distribution is also extremely weak.
For this reason, even when the surface pressure or surface shear stress is analysed directly, the small differences associated with lift are likely to be buried in the dominant drag-related components, making the origin of lift difficult to identify clearly.
Furthermore, in real fluid systems, the working fluid often exhibits non-Newtonian properties such as shear-thinning behaviour.
Such rheological effects can influence not only the magnitude of inertial lift but also, in some cases, its direction, through changes in the viscosity distribution and the stress field~\citep{wang2003lift,li2015dynamics,raoufi2019experimental,hu2020influence,hazra2021dynamics,yamashita2025focusing}.
Because the stress in non-Newtonian fluids responds nonlinearly to the local flow state, the relation between the surface-stress distribution and the underlying flow-field structure becomes more complicated, and it becomes even more difficult to interpret the origin of lift intuitively from the surface-stress distribution.

Motivated by these issues, the central question of this study is whether the origin of weak inertial lift in wall-bounded flows of Newtonian and non-Newtonian fluids can be interpreted not as a small difference in surface stress, but in relation to the internal structure of the flow field.
In this study, we consider the flow between eccentric rotating cylinders, in which this question arises particularly clearly. Fluid forces associated with this class of eccentric annular configurations are encountered in a wide variety of systems, including bearing flows \citep{szeri1995flow}, annular flows in drilling equipment \citep{volpi2024whirling}, and eccentric-rotation systems relevant to stirring and mixing \citep{watamura2025transition}. For this reason, many analytical and asymptotic expressions for lift have been derived in the limits of low Reynolds number, low eccentricity, or under the lubrication approximation~\citep{kamal1966separation,yamada1968flow,Ballal1976Flow_ag,DiPrima1972Flow_cp,brindley1979flow,kazakia1978flow,Brennen1976Flow_wd}, and substantial knowledge has therefore been accumulated regarding predictive expressions for the force itself.
On the other hand, attempts to interpret the mechanism of lift generation in connection with the flow structure remain limited.
Indeed, many previous studies that investigated inertial lift in this system numerically have suggested, on the basis of surface-force analysis, that the pressure contribution is dominant~\citep{SanAndres1984Flow_nl,Feng2007Orbital_ns,resell2025fluid}.
However, the asymmetry in surface pressure responsible for the lift is extremely weak compared with the pressure component contributing to drag, and the flow-field-side factors that generate this asymmetry remain unclear.
Furthermore, when non-Newtonian effects are taken into account, the problem becomes even more complicated.
In practical applications, slurries and polymer solutions are often used as working fluids in drilling fluids and lubricants~\cite{alderman1988rheological,berker1995effect,carrasco2010non}, and the influence of rheological effects such as shear-thinning behaviour and elasticity on inertial lift cannot be neglected.
Indeed, numerical studies of flow between eccentric rotating cylinders in shear-thinning fluids have reported that shear-thinning rheology modifies the lift and strengthens the tendency to push the inner cylinder toward the outer wall~\citep{podryabinkin2011moment}.
However, it is still not sufficiently understood how the non-uniform viscosity distribution leads to variations in lift through the resulting asymmetry of the surface-stress field.

Thus, the difficulty of causally identifying the origin of a force from the surface-stress distribution is not limited to eccentric rotating cylinders, but arises widely in fluid-mechanical problems such as aerodynamic forces around wings~\citep{birch2004force,otomo2021unsteady,seo2022improved}, hydrodynamic forces acting on swimming and flying bodies~\citep{sane2003aerodynamics}, and unsteady fluid forces on bodies accompanied by vortex shedding~\citep{barnes1983vortex, matsumoto1999vortex}.
For this reason, frameworks that evaluate and interpret fluid forces as volume integrals over the fluid domain, without directly relying on the stress distribution on the body surface, have long been developed (\citealp[Chapter 3]{saffman1992vortex};
\citet{Howe1995Force_gn,Noca1997Measuring_xl,Wu2007Integral_bg,Li2018Vortex_dp,menon2021initiation,Wang2022Vortex_hw,prakhar2025vortices,gao2025weighted}).
An important feature of these approaches is that fluid forces can be described in terms of physical quantities within the flow field, such as the velocity field, the vorticity field, and their spatial gradients.
Therefore, instead of directly tracing a small difference in surface stress, it becomes easier to identify which regions of the flow field contribute essentially to force generation and to discuss the relation between the force and the flow structure.
In particular, in the low-Reynolds-number regime of wall-bounded flow, reciprocal-theorem-type approaches based on the Lorentz reciprocal theorem have been established for the linear Stokes equations, in which the desired fluid force is expressed through an integral identity involving an auxiliary problem  (\citealp[Chapter 3]{happel1973low}).
More recently, \citet{Magnaudet2011Reciprocal_ue} extended reciprocal-theorem-based formulations to finite-Reynolds-number flows and presented a general expression in which the fluid force can be written as a volume integral in terms of physical quantities within the flow field even when the nonlinearity due to inertia is included.
In addition, \citet{Masoud2019Reciprocal_pf} suggested that this type of volume-integral expression can serve as a diagnostic tool for identifying the dominant contributions to the force and interpreting its generation mechanism by using flow-field data obtained from numerical simulations or experiments.
Nevertheless, to the best of the authors' knowledge, applications of these ideas to a full mechanism analysis of lift in wall-bounded flow remain limited.

Based on the above, the objective of this study is to present a framework for quantitatively extracting the causal relation between steady inertial lift in wall-bounded flow and the flow structures responsible for it, on the basis of flow-field data obtained from numerical simulations.
As the target problem, we consider the steady flow between eccentric rotating cylinders at low Reynolds numbers, and derive a volume-integral expression for lift for both a Newtonian fluid and a shear-thinning fluid, the latter representing one of the typical forms of non-Newtonian behaviour.
The key point of the present framework is that, starting from a force formulation based on the reciprocal theorem, the lateral force is expressed as a volume integral that contains the vortex-force contribution.
This expression has a structure common to force interpretations based on the vortex force that have been developed in aerodynamics and related fields, and provides a basis for describing and interpreting inertial lift in wall-bounded flow from the internal structure of the flow field.
In this study, we apply this framework to lift in eccentric rotating cylinders and address two questions.
First, we clarify what changes in the flow structure cause the sign reversal of lift induced by increasing eccentricity in a Newtonian fluid.
Second, focusing on the fact that shear-thinning behaviour can reverse the sign of lift even at the same eccentricity, we clarify what kind of modulation of the flow field causes this change.
By answering these questions, we show that the proposed framework is effective for analysing the mechanism of steady inertial lift in wall-bounded flow and for interpreting how shear-thinning behaviour affects this lift.

The paper is organised as follows.
In \S\ref{ssec:eq}, we first present the problem setting for eccentric rotating cylinders, the governing equations, and the constitutive relation for the shear-thinning fluid.
Then, in \S\ref{ssec:num}, we describe the numerical method and its validation.
Further, in \S\ref{ssec:frc}, based on the reciprocal theorem, we formulate lift as a volume-integral expression consisting of a vortex-force contribution and a viscous stress contribution.
In \S\ref{sec:result}, we present the results and discussion.
First, in \S\ref{ssec:flow}, we give an overview of the effects of shear-thinning behaviour on the flow field and the viscosity field.
Next, in \S\ref{ssec:drag}, we show the effect of shear-thinning behaviour on the drag.
Further, in \S\ref{ssec:lift}, we show the dependence of lift on the eccentricity and the power-law index, and clarify the sign distribution of lift and the conditions under which sign reversal occurs.
Based on these results, in \S\ref{ssec:mechanism_e}, we analyse the mechanism of eccentricity-induced lift reversal in the Newtonian fluid using the proposed lift-diagnostic framework.
Then, in \S\ref{ssec:mechanism_n}, we clarify the mechanism of lift reversal caused by shear-thinning behaviour.
Finally, concluding remarks are given in \S\ref{sec:con}.

\section{Methods}
\subsection {General formulation}
\label{ssec:eq}
\begin{figure}
  \centering
  \includegraphics[width=0.55\linewidth]{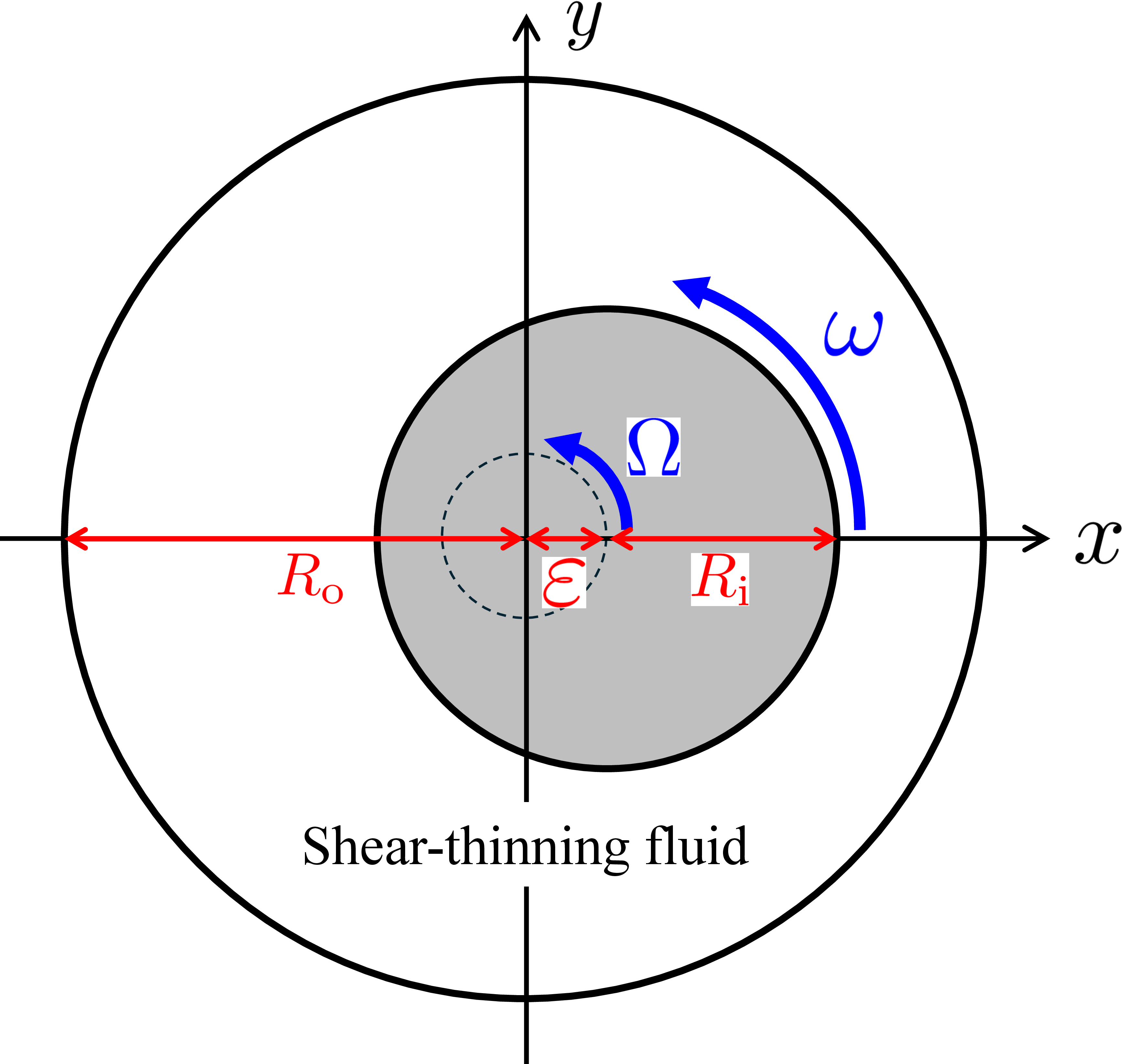}
  \caption{Schematic of the flow between eccentric rotating cylinders. The gap between the inner and outer cylinders is filled with a shear-thinning fluid. Here, $\varepsilon$ denotes the centre-to-centre distance between the cylinders, and $R_{\mathrm{i}}$ and $R_{\mathrm{o}}$ are the radii of the inner and outer cylinders, respectively. The angular velocities $\Omega$ and $\omega$ denote the orbital motion about the outer-cylinder centre and the rotation of the inner cylinder about its own centre, respectively. The origin of the Cartesian coordinate system $(x,y)$ is located at the centre of the outer cylinder.}
  \label{fig:fig1}
\end{figure}

To investigate the mechanism of hydrodynamic force generation, we consider the flow of a shear-thinning fluid between eccentric rotating cylinders and perform numerical simulations.
Hereafter, $(\cdot)^*$ denotes dimensional quantities.
The fluid fills the gap between an outer cylinder of radius $R_{\mathrm{o}}^*$ and an inner cylinder of radius $R_{\mathrm{i}}^*$, and the flow is driven by the rotation and orbital motion of the inner cylinder.
A schematic of the configuration, together with the Cartesian coordinate system, is shown in figure~\ref{fig:fig1}.
The centres of the outer and inner cylinders are located at the origin and at $(\varepsilon^*,0)$, respectively.
The eccentricity is defined as $e=\varepsilon^*/(R^*_{\mathrm{o}}-R^*_{\mathrm{i}})$.
The outer cylinder is stationary, while the inner cylinder rotates about its own centre with angular velocity $\omega^*$ and revolves about the centre of the outer cylinder with angular velocity $\Omega^*$.
In the present study, the power-law model is adopted as the constitutive equation for the shear-thinning fluid:
\begin{equation}
 \mu^*(\dot{\gamma}^*) = K^* \dot{\gamma}^{*n-1},
 \label{eq:power_law}
\end{equation}
where $\mu^*(\dot{\gamma}^*)$ is the viscosity, $K^*$ is the consistency coefficient, $\dot{\gamma}^* = \sqrt{2\bm{S}^*\colon\bm{S}^*}$ is the strain rate, and $\bm{S}^*=(\nabla\bm{u}^*+(\nabla\bm{u}^*)^{\mathrm{T}})/2$ is the strain-rate tensor.
The power-law index $n$ characterises the dependence of the viscosity $\mu^*$ on the strain rate $\dot{\gamma}^*$, with $n<1$ corresponding to shear-thinning behaviour.
We note that more elaborate models, such as the Cross one~\citep{cross1965rheology} and the Carreau--Bird one~\citep{carreau1972rheological}, are also widely used for shear-thinning viscosity.
However, in flows such as the present one, in which the near-wall high-shear region is dominant, these models effectively reduce to power-law behaviour.
For this reason, we use equation~\eqref{eq:power_law} in the present study.
Hereafter, all variables are non-dimensionalised using the reference quantities
\begin{equation}
  \begin{gathered}
    U^*=\max(R^*_{\mathrm{i}}\omega^*, (R^*_{\mathrm{o}}-R^*_{\mathrm{i}})\Omega^* ),\quad
    L^*=R^*_{\mathrm{o}}-R^*_{\mathrm{i}},  \\
    t^*=\frac{L^*}{U^*},\quad
    p^*=\rho^* U^{*2},\quad
    \mu^*=K^*\left(\frac{U^*}{L^*}\right)^{n-1}.
  \end{gathered}
\end{equation}
The Reynolds number is defined as $\operatorname{Re}=\rho^* U^* L^*/\mu^*$.
For simplicity, the same symbols are used hereafter for the dimensionless variables.

The governing equations for the fluid are the incompressibility condition and the Navier--Stokes equations in a moving frame rotating about the origin with angular velocity $\bm{\Omega}$:
\begin{align}
  &\nabla \cdot \bm{u} = 0, \label{eq:cnt}\\
  & \frac{\partial \bm{u}}{\partial t} + (\bm{u}^{\prime}\cdot\nabla)\bm{u} + \bm{\Omega} \times \bm{u}= -\nabla p^{\prime} + \frac{1}{\operatorname{Re}} \nabla \cdot \bm{\tau},
  \label{eq:ns}
\end{align}
where $\bm{u}$ is the velocity in the inertial frame, and $\bm{u}^{\prime}=\bm{u}-\bm{\Omega}\times\bm{x}$ is the velocity relative to the moving frame.
$\bm{\Omega}=\Omega \bm{e}_z$ is the angular-velocity vector of the moving frame, and $\bm{x}$ is the position vector measured from the origin.
$p^{\prime}=p+p_{\mathrm{rot}}$ is the modified pressure, defined as the sum of the pressure $p$ in the inertial frame and the correction potential $p_{\mathrm{rot}}$ associated with the rotation of the coordinate system.
$\bm{\tau}=2 \mu(\dot{\gamma}) \bm{S}$ is the shear-stress tensor.
No-slip boundary conditions are imposed on the inner-cylinder surface $S_{\mathrm{i}}$ and the outer-cylinder surface $S_{\mathrm{o}}$:
\begin{alignat}{2}
  \bm{u} &= \varepsilon \Omega \bm{e}_y
          + \bm{\omega} \times \bigl(\bm{x}-\bm{x}_{\mathrm{i}}\bigr)
  \quad && \text{on } S_{\mathrm{i}}, \label{eq:bci}\\
  \bm{u} &= \bm{0}
  \quad && \text{on } S_{\mathrm{o}},  \label{eq:bco}
\end{alignat}
where $\varepsilon \Omega \bm{e}_y$ denotes the orbital velocity of the inner cylinder about the centre of the outer cylinder, and $\bm{\omega} \times\bigl(\bm{x}-\bm{x}_{\mathrm{i}}\bigr)$ denotes the rotational velocity about the centre of the inner cylinder.
Here, $\bm{\omega} = \omega \bm{e}_z$ is the angular-velocity vector of the inner-cylinder rotation, and $\bm{x}_{\mathrm{i}}$ is the position vector of the centre of the inner cylinder measured from the origin.

In the present study, of the hydrodynamic force $\bm{F}$ acting on the inner cylinder, the component perpendicular to the orbital direction (the $x$-component) is defined as the lift $F_x$, while the component parallel to the orbital direction (the $y$-component) is defined as the drag $F_y$.
When $\bm{F}$ is evaluated by a surface integral, we use the pressure $p$ defined in the inertial frame and omit the contribution arising from the correction potential $p_{\mathrm{rot}}$, following \citet{Feng2007Orbital_ns,kazakia1978flow}.
This is because $p_{\mathrm{rot}}$ does not represent the surface force exerted on the inner cylinder by the surrounding fluid, but rather contributes to the centripetal force required to maintain the orbital motion.
Accordingly, in the present study, $\bm{F}$ is evaluated as
\begin{equation}
 \bm{F} = -\int_{S_{\mathrm{i}}} \bm{n} \cdot \left( -p\bm{I} + \operatorname{Re}^{-1}\bm{\tau} \right) ~ \mathrm{d} S
 \label{eq:frc_s},
\end{equation}
where $\bm{n}$ is the unit normal vector directed from the fluid into the interior of the inner cylinder.

\subsection {Numerical simulation method}
\label{ssec:num}
We discretised the governing equations in the bipolar coordinate system $(\xi,\eta)$ and solved them numerically using a second-order finite-difference method.
The computational grid used in this study is boundary-fitted to both the outer and inner cylinders (see figure~\ref{fig:app1} in \hyperref[sec:appA]{Appendix~\ref{sec:appA}}).
Technical details of the discretisation of equations~\eqref{eq:cnt}--\eqref{eq:frc_s} are given in \hyperref[sec:appA]{Appendix~\ref{sec:appA}}.
The number of grid points was set to $N_\xi \times N_\eta = 128 \times 256$, and a non-uniform grid was employed so that the grid was refined near both the outer and inner cylinders.

For time integration, we used a second-order semi-implicit scheme based on the simplified marker-and-cell method~\citep{amsden1970simplified}.
Technical details of the solution procedure for the pressure Poisson equation arising in the SMAC method are given in \hyperref[sec:appB]{Appendix~\ref{sec:appB}}.
Specifically, the convective term was discretised using the explicit second-order Adams-Bashforth method, whereas the viscous term was discretised using the implicit second-order Crank-Nicolson method, and the unsteady equations were advanced in time until a steady solution was obtained.
Convergence was assessed on the basis of the kinetic-energy balance over the entire computational domain.
More specifically, after sufficient time advancement, we confirmed that the mechanical energy input rate, calculated by a surface integral over the inner-cylinder surface, agreed with the kinetic-energy dissipation rate evaluated over the entire computational domain.
In fact, the relative error between them was less than $6.0 \times 10^{-5}\%$ over all parameter ranges discussed below.
These results confirm that each computation had converged and reached a steady state from the viewpoint of the kinetic-energy balance.

The power-law model~\eqref{eq:power_law} adopted in this study has the well-known difficulty that the viscosity diverges in the low-shear limit $\dot{\gamma}\to 0$ (\citealp[Chapter 1]{chhabra2023bubbles}).
To avoid numerical instability caused by this difficulty, \citet{zhang2023drag} introduced an upper bound $\mu_{\max}$ for the viscosity and defined it as $\mu = \min(\dot{\gamma}^{\,n-1},\, \mu_{\max})$.
Following \citet{zhang2023drag}, we likewise imposed an upper limit on the viscosity and set $\mu_{\max}=1\times10^{3}$.
To assess this choice, we carried out a sensitivity analysis of the lift $F_x$ with respect to $\mu_{\max}$.
Specifically, for four representative parameter sets corresponding to the extreme values of $e$ and $\Omega/\omega$ within the parameter range described later in this subsection, we considered the strongest shear-thinning case, $n=0.2$, and systematically varied the upper bound over the range $\mu_{\max}\in[1,10^4]$ to calculate the lift.
As a result, the variation in the lift became weaker as $\mu_{\max}$ increased and reached a plateau around $\mu_{\max}=10^2$.
The absolute error in the lift between $\mu_{\max}=4\times10^2$ and $\mu_{\max}=10^3$ was less than $0.03\%$ for all tested parameter sets.
Therefore, the value $\mu_{\max}=10^3$ used in this study is sufficiently large, and the flow behaves essentially as that of a pure power-law fluid within the parameter range considered here.

\begin{figure}
  \centering
  \begin{minipage}{0.48\linewidth}
    \centering
    \begin{overpic}[width=\linewidth]{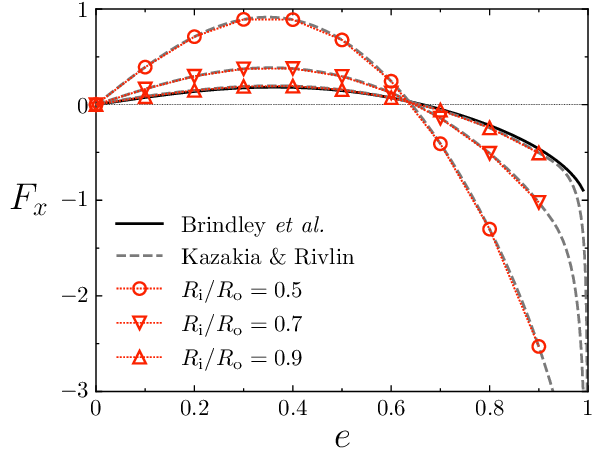}
      \put(0,74){\large (a)}
    \end{overpic}
  \end{minipage}
  \hspace{3pt}
  \begin{minipage}{0.48\linewidth}
    \centering
    \begin{overpic}[width=\linewidth]{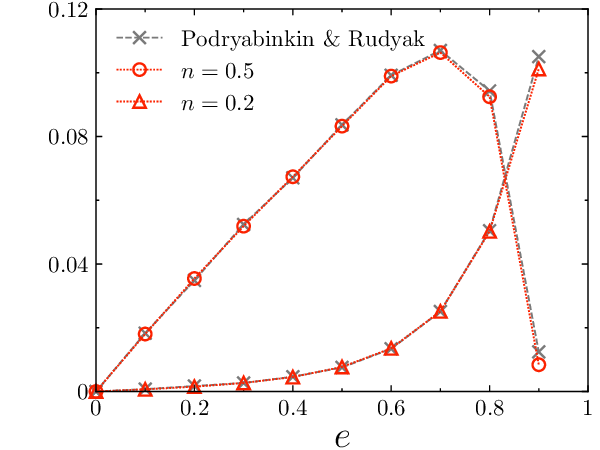}
      \put(0,74){\large (b)}
    \end{overpic}
  \end{minipage}
  \caption{Dependence of the lift force $F_x$ on the eccentricity $e$ in an eccentric rotating-cylinder system. Red symbols indicate the results of the present numerical simulations. (a) Newtonian fluid case ($n = 1.0$) at $\operatorname{Re} = 1$ and $\Omega/\omega = 1$. The solid and dashed lines represent the analytical solutions of \citet{brindley1979flow} and \citet{kazakia1978flow}, respectively. (b) Shear-thinning power-law fluid cases at $\operatorname{Re} = 100$ with $n = 0.5$ and $0.2$, where the inner cylinder rotates without orbital motion ($\Omega = 0$). Grey symbols indicate the numerical results reported by \mbox{\citet{podryabinkin2011moment}}.}
  \label{fig:val}
\end{figure}

To confirm the reliability of the lift obtained in the present computations, we compared our results with analytical and numerical results reported in the literature.
Figure~\ref{fig:val}(a) compares the lift for a Newtonian fluid ($n=1$) at $\operatorname{Re}=1$ and $\Omega/\omega=1$.
The red markers denote the present numerical results, the black solid line denotes the asymptotic solution of \citet{brindley1979flow} based on the lubrication approximation (the limit $R_{\mathrm{i}}/R_{\mathrm{o}}\to1$), and the gray dashed line denotes the approximate solution of \citet{kazakia1978flow}, which includes the first-order correction due to inertia.
For the radius ratio $R_{\mathrm{i}}/R_{\mathrm{o}}=0.5$, the present lift agrees well with the approximate solution of \citet{kazakia1978flow}, and the relative error was less than $0.12\%$ for all eccentricities.
As the radius ratio is increased systematically, the present numerical results approach the lubrication solution of \citet{brindley1979flow} in a consistent manner.
Figure~\ref{fig:val}(b) compares the present numerical results (red markers) with the numerical results of \citet{podryabinkin2011moment} (gray markers) for the power-law indices $n=0.2$ and $0.5$ at $\operatorname{Re}=100$ and $R_{\mathrm{i}}/R_{\mathrm{o}}=0.5$.
For shear-thinning fluids, few comparable previous studies are available, and here we refer to the numerical results reported by \citet{podryabinkin2011moment} for the case in which the inner cylinder undergoes only self-rotation ($\Omega=0$).
The present results exhibit good agreement regardless of the eccentricity and the power-law index.
These comparisons confirm that the numerical method used in this study is sufficiently reliable for evaluating lift in power-law fluids.

In the present numerical simulations, the Reynolds number $\operatorname{Re}=1$ and the radius ratio $R_{\mathrm{i}}/R_{\mathrm{o}}=0.5$ were fixed, and the three parameters $e$, $\Omega/\omega$, and $n$ were varied systematically.
The eccentricity was varied from $e=0.1$ to $0.9$.
The rotation-rate ratio was varied over the range $\Omega/\omega=0.02$ to $2$ so as to cover a wide range from rotation-dominated to orbital-dominated conditions.
The power-law index was varied from $n=0.2$ to $1.0$ so that the continuous transition from strong shear-thinning behaviour to Newtonian behaviour could be examined.

\subsection {Formulation of lift force}
\label{ssec:frc}
In this section, we derive the lift formulation that constitutes the core of the present paper.
Specifically, following the procedure used in derivations based on the Lorentz reciprocal theorem (\citealp[Chapter 3]{happel1973low}) and its related applications~\citep{Magnaudet2011Reciprocal_ue,Masoud2019Reciprocal_pf}, we transform the lift expression defined by the surface integral~\eqref{eq:frc_s} into a volume-integral expression over the fluid domain.
The resulting expression is similar to formulations based on the Lamb vector $\bm{l}=\bm{u}\times\bm{\omega}$ (or vortex force), which have been used to discuss the mechanisms of lift generation and hydrodynamic-force generation in relation to the vorticity field (\citealp[Chapter  3]{saffman1992vortex}; \citealp[Chapter 11]{wu2006vorticity}).
To derive this formulation, we first rewrite the governing equations~\eqref{eq:cnt}-\eqref{eq:bco} in a frame rotating about the origin with angular velocity $\bm{\Omega}$, corresponding to the orbital motion of the inner cylinder.
In this frame, centrifugal and Coriolis acceleration terms appear.
However, in the two-dimensional system considered here, both can be written as gradients of scalar potentials.
They can therefore be absorbed into the correction potential gradient $\nabla p_{\mathrm{rot}}$ in equation~\eqref{eq:ns}.
Following \citet{Feng2007Orbital_ns,kazakia1978flow}, we subtract these potential-gradient terms from the equation of motion and focus only on the hydrodynamic force based on the pressure $p$ in the inertial frame.
We further restrict attention to steady lift and neglect temporal variations in the flow field.
Then, the governing equations and boundary conditions become
\begin{subequations}
\begin{align}
 \nabla \cdot \bm{u}^{\prime} &= 0,
 \label{eq:cnt2}\\ 
 \nabla \cdot \bm{\sigma}  &= \nabla \left( \frac{|\bm{u}^{\prime}|^2}{2} \right) -  \bm{l}^{\prime},
 \label{eq:ns2}\\
 \bm{\sigma} &= -p\bm{I} + \operatorname{Re}^{-1} 2\mu\bm{S}, \quad \mu = \dot{\gamma}^{n-1},
 \label{eq:sigma}\\
 \bm{u}^{\prime} &=
 \begin{cases}
 \bm{\omega} \times (\bm{x}-\bm{x}_{\mathrm{i}}) & \text{on } S_{\mathrm{i}},\\
 -\bm{\Omega} \times \bm{x}  & \text{on } S_{\mathrm{o}},
 \end{cases}
 \label{eq:bc2}
\end{align}
\end{subequations}
where $\bm{u}^{\prime}=\bm{u} - \bm{\Omega} \times \bm{x}$ is the velocity relative to the rotating frame, $\bm{l}^{\prime}=\bm{u}^{\prime} \times \bm{\omega}^{\prime}$ is the Lamb vector (or vortex force (\citet[Chapter I]{von1935general};\citet[Chapter 3]{saffman1992vortex})), and $\bm{\omega}^{\prime}=\nabla \times \bm{u}^{\prime}$ is the relative vorticity.
Next, following the standard derivation of the reciprocal theorem, we introduce an auxiliary field $(\hat{\bm{u}},\hat{\bm{\sigma}})$.
The auxiliary field is defined as the steady, incompressible Newtonian Stokes flow generated by an inner cylinder translating in the $\bm{e}_x$ direction:
\begin{subequations}
\begin{align}
 \nabla \cdot \hat{\bm{u}} &= 0, \label{eq:auxx_cnt}\\
 \nabla \cdot \hat{\bm{\sigma}} &= 0, \label{eq:auxx_mom}\\
 \hat{\bm{u}} &=
 \begin{cases}
 \bm{e}_x & \text{on } S_{\mathrm{i}}, \\
 \bm{0}   & \text{on } S_{\mathrm{o}}.
 \end{cases}
 \label{eq:auxx_bc}
\end{align}
\end{subequations}
We then multiply \eqref{eq:auxx_mom} by $\bm{u}^{\prime}$ and \eqref{eq:ns2} by $\hat{\bm{u}}$, subtract the latter from the former, and integrate the result over the fluid domain $V$.
Because $\nabla\cdot\hat{\bm{\sigma}}=\bm{0}$, the volume contribution arising from \eqref{eq:auxx_mom} vanishes identically.
After further rearrangement using the divergence theorem and the incompressibility condition, we obtain
\begin{equation}
\begin{aligned}
-\int_{S_\mathrm{i} \cup S_\mathrm{o}} \bm{n} \cdot \bm{\sigma} \cdot \hat{\bm{u}}\, \mathrm{d}S
+\int_V \bm{\sigma} : \nabla \hat{\bm{u}} \, \mathrm{d}V 
=
-\int_{S_\mathrm{i} \cup S_\mathrm{o}}
(\bm{n}\cdot\hat{\bm{u}})
\left(\frac{\lvert \bm{u}^{\prime} \rvert^2}{2}\right)\, \mathrm{d}S
+\int_V \hat{\bm{u}} \cdot \bm{l}^{\prime} \, \mathrm{d}V .
\label{eq:lrt1}
\end{aligned}
\end{equation}
We now simplify each term in equation~\eqref{eq:lrt1} in turn.
In the first term on the left-hand side, the contribution from $S_{\mathrm{o}}$ vanishes because $\hat{\bm{u}}=\bm{0}$ there by the boundary condition~\eqref{eq:auxx_bc}.
On the other hand, on $S_{\mathrm{i}}$, we have $\hat{\bm{u}}=\bm{e}_x$, so this term coincides with the definition of the lift in \eqref{eq:frc_s}.
In the second term on the left-hand side, the pressure contribution included in the stress tensor $\bm{\sigma}$ vanishes by incompressibility, since it can be rewritten as $-p \bm{I} \colon \nabla \hat{\bm{u}} = -p \nabla \cdot \hat{\bm{u}}$.
In the first term on the right-hand side, the contribution from $S_{\mathrm{o}}$ is zero because $\hat{\bm{u}}=\bm{0}$ there.
On $S_{\mathrm{i}}$, from the boundary condition~\eqref{eq:bc2}, $\lvert \bm{u}^{\prime}\rvert^2/2$ is constant and can therefore be taken outside the integral.
The remaining integral, $\int_{S_{\mathrm{i}}}\hat{\bm{u}}\cdot\bm{n}\,\mathrm{d}S$, vanishes by the divergence theorem and the incompressibility condition, so the first term on the right-hand side also disappears.

Hence, equation~\eqref{eq:lrt1} becomes
\begin{equation}
 F_x = \underbrace{\int_{V} \hat{\bm{u}} \cdot \bm{l}^{\prime} \, \mathrm{d}V}_{F_x^{(l)}}
 + \underbrace{\frac{1}{\operatorname{Re}} \int_{V} (- \bm{\tau}) \colon \hat{\bm{S}} \, \mathrm{d}V}_{F_x^{(\tau)}},
 \label{eq:frc_v}
\end{equation}
where $F_x^{(l)}$ denotes the vortex-force contribution associated with the Lamb vector $\bm{l}^{\prime}=\bm{u}^{\prime}\times\bm{\omega}^{\prime}$, while $F_x^{(\tau)}$ denotes the viscous stress contribution arising from the viscous stress field.
Hereafter, we refer to these as the vortex-force contribution and the viscous stress contribution, respectively.
Note that in the Newtonian case $n = 1$, although the integral kernel of the shear-stress component $F_x^{(\tau)}$ spatially varies, $F_x^{(\tau)} = 0$ holds because
\begin{equation}
\begin{aligned}
\int_{V} (- \bm{\tau}) \colon\hat{\bm{S}} ~ \mathrm{d} V
&=
-\int_{S_{\rm i} \cup S_{\rm o}} {\bm n}\cdot \hat{\bm\sigma} \cdot {\bm u} ~ \mathrm{d} S \\
&=
-\varepsilon \Omega
\underbrace{\int_{S_{\rm i}} {\bm n}\cdot \hat{\bm\sigma} \cdot \bm{e}_y ~ \mathrm{d} S}_{=0}
-\omega
\underbrace{\int_{S_{\rm i}} {\bm n}\cdot \hat{\bm\sigma} \cdot (\bm{e}_z \times (\bm{x}-\bm{x}_{\mathrm{i}})) ~ \mathrm{d} S}_{=0}
=0
\label{eq:frc_v_n1}
\end{aligned}
\end{equation}
owing to the fore-aft symmetry in the auxiliary field.

An advantage of equation~\eqref{eq:frc_v} is that it expresses the lift in this decomposed form as a volume integral.
By substituting into this equation the velocity field and its spatial derivatives obtained from numerical simulations or experiments, the lift can be evaluated separately in terms of the vortex-force contribution and the viscous stress contribution.
Furthermore, by examining the spatial distributions of the integrands of these contributions, this formulation provides a basis for discussing where in the fluid domain the lift is generated and how it is related to the flow structure.
In the following sections, \S~\ref{ssec:mechanism_e} and \ref{ssec:mechanism_n}, we apply equation~\eqref{eq:frc_v} to the actual flow fields and analyse the physical mechanism of lift.

\section{Results and Discussion}
\label{sec:result}
\subsection{Flow and viscosity fields}
\label{ssec:flow}

\begin{figure}
  \centering
  \vspace{10pt}
  \begin{minipage}{0.98\linewidth}
    \centering
    \begin{overpic}[width=\linewidth]{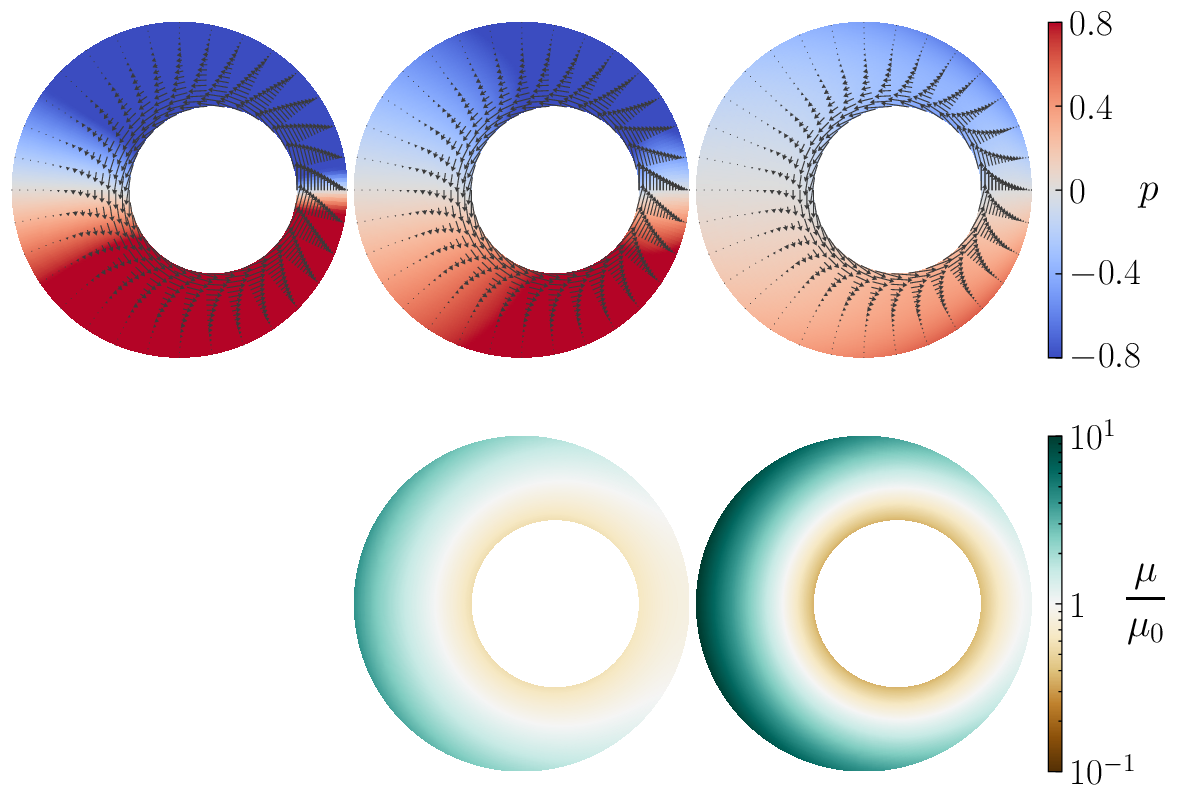}
      \put(12, 32){\large $n=1$}
      \put(39.5,   32){\large $n=0.6$}
      \put(69,   32){\large $n=0.4$}
      \put(0,67){\large (a) $~~\Omega/\omega = 0.02$}
    \end{overpic}
  \vspace{10pt}
  \end{minipage}
  \begin{minipage}{0.98\linewidth}
    \centering
    \begin{overpic}[width=\linewidth]{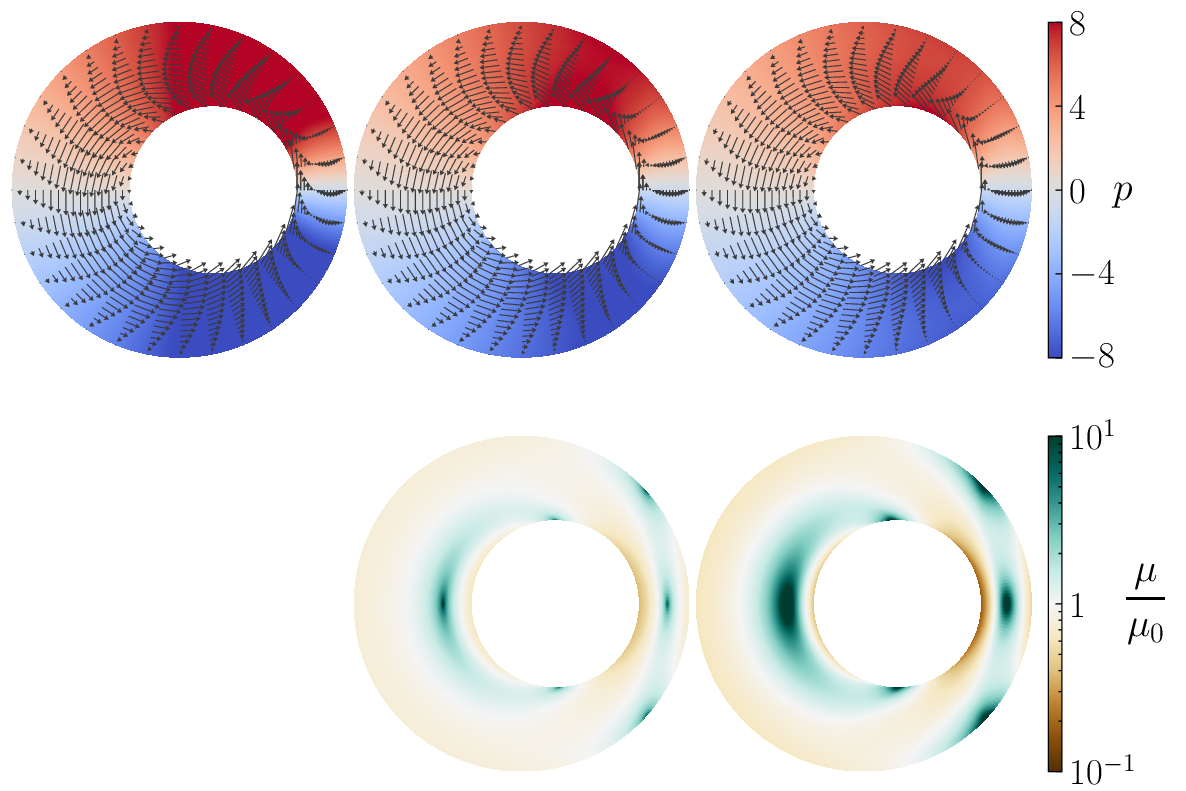}
      \put(12, 32){\large $n=1$}
      \put(39.5,   32){\large $n=0.6$}
      \put(69,   32){\large $n=0.4$}
      \put(0,67){\large (b) $~~\Omega/\omega = 2$}
      \put(6,6){\includegraphics[width=0.1\linewidth]{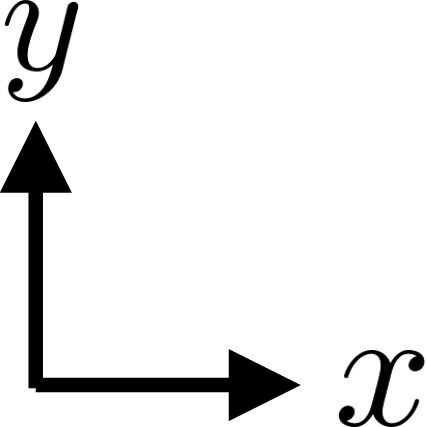}}
    \end{overpic}
  \end{minipage}
  \caption{
    Flow fields at eccentricity $e = 0.4$ for several power-law indices $n$:
    $n=1$, $0.6$, and $0.4$.
    Panels (a) and (b) correspond to the rotation-rate ratios
    $\Omega/\omega = 0.02$ and $2$, respectively.
    In each panel, the top row shows colour contours of the pressure $p$
    with black arrows indicating the velocity vectors,
    while the bottom row shows the spatial distribution of the viscosity ratio
    $\mu/\mu_0$, where $\mu_0$ is the viscosity of the Newtonian fluid.
  }
  \label{fig:fig3}
\end{figure}

Before discussing the effect of shear-thinning behaviour on inertial lift, we first give an overview of how the power-law index $n$ modifies the flow field.
Figure~\ref{fig:fig3} shows the flow fields at eccentricity $e=0.4$ for two rotation-rate ratios, (a) $\Omega/\omega=0.02$ and (b) $\Omega/\omega=2$, and for three power-law indices, $n=1.0$, $0.6$, and $0.4$.
In the upper row of each panel, the spatial distributions of the pressure $p$ and the velocity vectors normalised by the maximum flow speed in each case are shown, while the lower row shows the spatial distribution of the viscosity ratio $\mu/\mu_0$, normalised by the Newtonian viscosity $\mu_0$.
We first focus on the velocity and pressure fields for the Newtonian fluid $(n=1)$ in figure~\ref{fig:fig3}(a).
The inner-cylinder rotation induces an azimuthal flow, and this flow enters the narrow-gap region, producing a positive-pressure region in the lower part of the gap $(y<0)$ and a negative-pressure region in the upper part $(y>0)$.
In contrast, for the shear-thinning fluids $(n=0.6,\,0.4)$, the rotation-driven azimuthal flow becomes weaker, and the magnitudes of both the positive and negative pressures decrease.
The viscosity also decreases markedly near the inner cylinder, where the shear rate is high.
Such viscosity reduction in the high-shear region is considered to weaken the rotation-driven flow, which in turn reduces the magnitudes of both the positive and negative pressures.
Next, we consider the velocity and pressure fields for the Newtonian fluid $(n=1)$ in figure~\ref{fig:fig3}(b).
Because of the orbital motion of the eccentric inner cylinder, the fluid is pushed in the upper part of the gap $(y>0)$, resulting in a high-pressure region there, whereas a low-pressure region forms in the lower part $(y<0)$.
This pressure difference drives pressure-driven flow in both the narrow-gap and wide-gap regions.
For the shear-thinning fluids $(n=0.6,\,0.4)$, it is again seen, as in figure~\ref{fig:fig3}(a), that both the positive and negative pressures decrease as $n$ decreases.
At the same time, the viscosity field in the lower row shows a pronounced reduction in viscosity very close to both the inner- and outer-cylinder surfaces.
This indicates that the viscous resistance is reduced in the high-shear regions near the walls and, as a result, the positive and negative pressures required to drive the flow also become smaller.
In the following sections, we examine how these changes in the flow field are related to the hydrodynamic forces.

\subsection{Drag forces}
\label{ssec:drag}

\begin{figure}
  \vspace{10pt}
  \centering
  \begin{minipage}{0.48\linewidth}
    \centering
    \begin{overpic}[width=\linewidth]{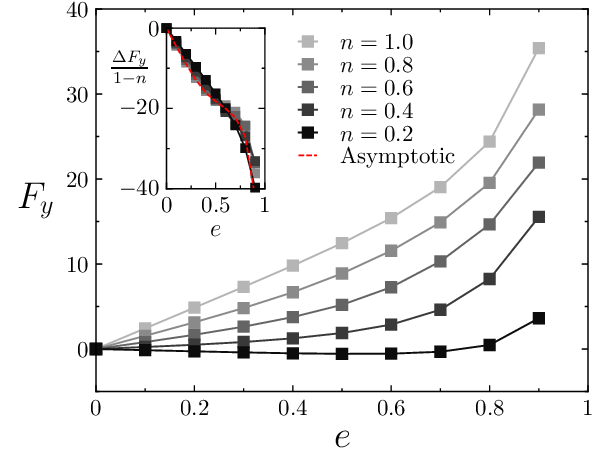}
      \put(5,77){\large (a) $~~\Omega / \omega = 0.02$}
    \end{overpic}
  \end{minipage}
  \hspace{3pt}
  \begin{minipage}{0.48\linewidth}
    \centering
    \begin{overpic}[width=\linewidth]{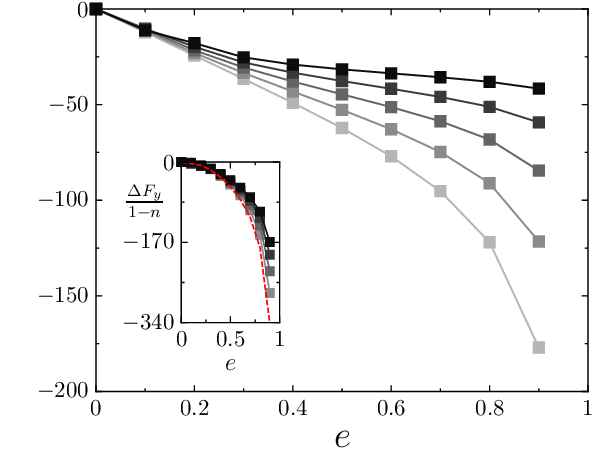}
      \put(5,77){\large (b) $~~\Omega / \omega = 2$}
    \end{overpic}
  \end{minipage}
  \caption{Dependence of the drag force $F_y$ on the eccentricity $e$ in a power-law fluid for two different rotation-rate ratios: (a) rotation-dominated regime $(\Omega / \omega = 0.02)$ and (b) orbital-dominated regime $(\Omega / \omega = 2)$. Each curve corresponds to a different power-law index $n$.  The inset shows the normalised variation $\Delta F_y / (1-n)$, where $\Delta F_y = F_y - F_{y,0}$ and $F_{y,0}$ denotes the drag force in a Newtonian fluid. The red dashed curve indicates the leading-order asymptotic expression in integral form, derived in the limit $1-n \ll 1$ (see~\hyperref[sec:app_asymp]{Appendix~\ref*{sec:app_asymp}}).}
  \label{fig:fig4}
\end{figure}

In this section, we examine the behaviour of the drag force $F_y$, for which the effect of shear-thinning behaviour appears relatively clearly among the hydrodynamic forces.
Figure~\ref{fig:fig4} shows $F_y$ as a function of $e$ for several values of $n$ at the rotation-rate ratios (a) $\Omega/\omega = 0.02$ and (b) $\Omega/\omega = 2$.
We first focus on the Newtonian fluid $(n=1)$.
Under the rotation-dominated condition in figure~\ref{fig:fig4}(a), $F_y$ is positive and increases monotonically with increasing $e$.
By contrast, under the orbital-dominated condition in figure~\ref{fig:fig4}(b), $F_y$ is negative and decreases monotonically as $e$ increases.
These signs are consistent with the directions of the force expected from the oppositely signed pressure fields forming above and below the gap, as shown in figure~\ref{fig:fig3}.
The monotonic dependence of $F_y$ on $e$ also indicates that the magnitudes of the positive and negative pressures change monotonically with increasing $e$.
Next, for the shear-thinning fluids $(n<1)$, the magnitude of $F_y$ decreases monotonically as $n$ decreases in both figure~\ref{fig:fig4}(a) and figure~\ref{fig:fig4}(b).
Therefore, although the sign of $F_y$ differs between the two conditions, the effect of shear-thinning behaviour consistently acts to reduce the magnitude of the drag force.
This is again consistent with figure~\ref{fig:fig3}, which shows that the magnitudes of both the positive and negative pressures forming above and below the inner cylinder decrease as $n$ decreases.

From these results, the effect of shear-thinning behaviour on the drag force $F_y$ can be understood in a relatively simple and intuitive manner.
That is, the sign of $F_y$ is determined by the pressure asymmetry between the upper and lower sides of the inner cylinder, and its variations with respect to $e$ and $n$ can be interpreted as monotonic changes in this pressure difference.
Indeed, each inset shows the quantity $\Delta F_y/(1-n)$, where $\Delta F_y=F_y-F_{y,0}$ and $F_{y,0}$ denotes the Newtonian value, and the results collapse approximately onto a single curve, indicating an almost linear response with respect to $n$.
In addition, when the prediction from the asymptotic analysis derived under the assumption $1-n\ll1$ (Appendix~\S\ref{sec:app_asymp}) is superposed as a red dashed line, it agrees well with the numerical results at least up to moderate eccentricities.
Thus, the fact that $\Delta F_y/(1-n)$ is organised approximately as a function of $e$ indicates that, to leading order, the shear-thinning-induced change in drag is proportional to $(1-n)$.

\subsection{lift forces}
\label{ssec:lift}

\begin{figure}
  \vspace{10pt}
  \centering
  \begin{overpic}[width=0.99\linewidth]{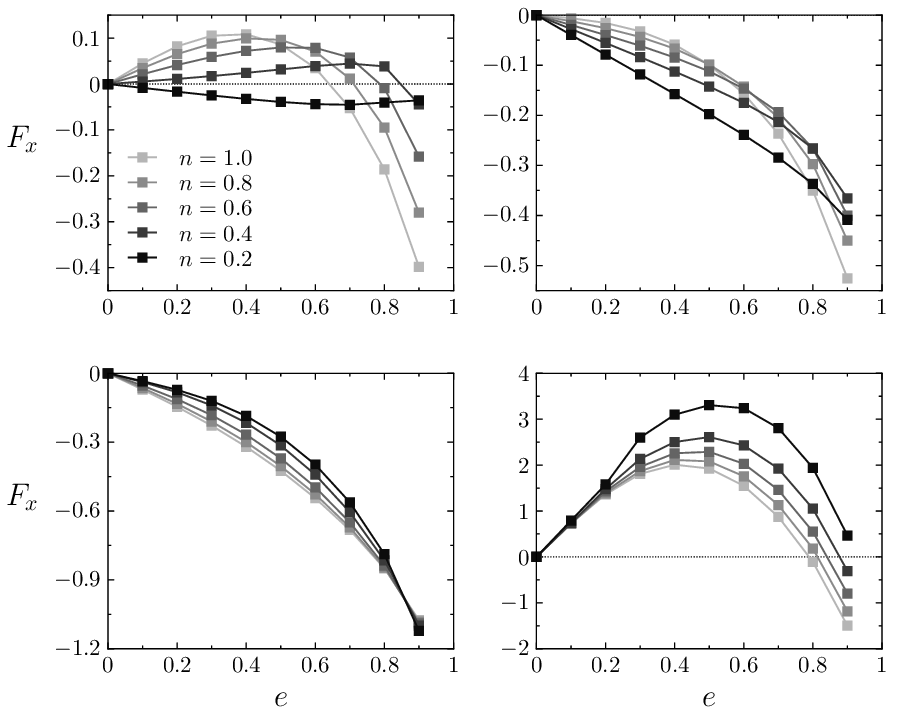}
      \put( 3,    81){\large (a) $~~\Omega / \omega = 0.02$}
      \put(53,    81){\large (b) $~~\Omega / \omega = 0.1$ }
      \put( 3,    41){\large (c) $~~\Omega / \omega = 0.4$   }
      \put(53,    41){\large (d) $~~\Omega / \omega = 2$   }
  \end{overpic}
  \caption{Dependence of the lift force $F_x$ on the eccentricity $e$ in a power-law fluid for three different rotation-rate ratios: (a) $\Omega / \omega = 0.02$, (b) $\Omega / \omega = 0.1$, (c) $\Omega / \omega = 0.4$ and (d) $\Omega / \omega = 2$. Each curve corresponds to a different power-law index $n$.}
  \label{fig:fig5}
\end{figure}
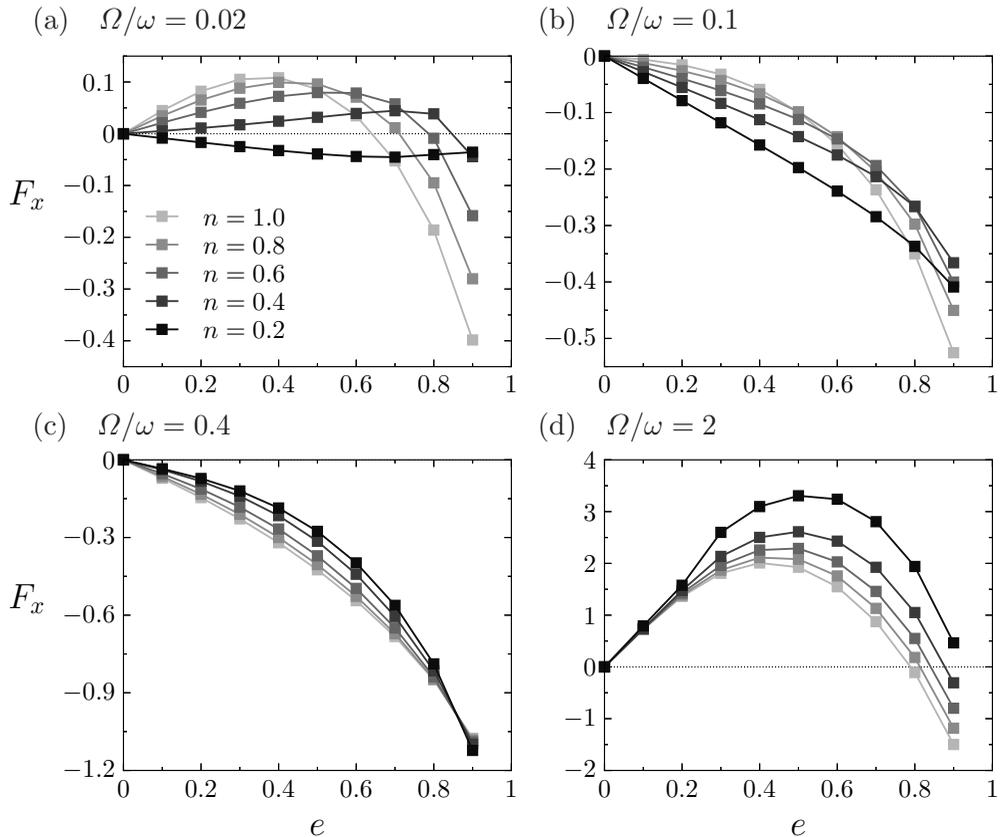

As shown in the previous section, the effect of shear-thinning behaviour on the drag force $F_y$ can be understood relatively clearly from its correspondence with the pressure distribution.
By contrast, the lift force $F_x$ exhibits more complex behaviour, because it is determined by the interaction between two nonlinear effects: fluid inertia and shear-thinning behaviour.
Figure~\ref{fig:fig5} shows the dependence of $F_x$ on $e$ for several power-law indices $n$ at the rotation-rate ratios $\Omega/\omega=$ (a) $0.02$, (b) $0.1$, (c) $0.4$, and (d) $2$.
We first focus on $F_x$ for the Newtonian fluid $(n=1)$.
In the rotation-dominated case shown in figure~\ref{fig:fig5}(a), $F_x$ varies non-monotonically with $e$, and there exists a zero-lift eccentricity within the range $0<e<1$.
As the rotation-rate ratio increases, $F_x$ becomes negative over the entire range in figure~\ref{fig:fig5}(b,c), and decreases monotonically with increasing $e$.
Furthermore, in figure~\ref{fig:fig5}(d), where the orbital effect is strong, a zero-lift eccentricity again appears within $0<e<1$.
Such a lift behaviour is non-trivial and cannot be understood readily from flow-field visualisation alone.
Indeed, in the pressure fields shown in figure~\ref{fig:fig3}, a clear asymmetry is observed mainly in the vertical direction, whereas the asymmetry in the horizontal direction is not evident.
Next, we describe how $F_x$ changes as $n$ decreases.
Here, we focus in particular on the variation of the zero-lift eccentricity.
In figure~\ref{fig:fig5}(a), as $n$ decreases, the location giving the maximum lift shifts toward higher $e$, and accordingly the eccentricity satisfying $F_x=0$ also shifts higher.
By contrast, in figure~\ref{fig:fig5}(d), $F_x$ increases over a wide range of $e$ as $n$ decreases, and, as a result, the zero-lift eccentricity likewise shifts higher.
On the other hand, under the conditions shown in figure~\ref{fig:fig5}(b,c), where no zero-lift eccentricity exists, the variation of $F_x$ with changing $n$ is relatively small.
Although not shown in the figure, we have confirmed that the same tendency is also observed at rotation-rate ratios between (a) and (b), and between (c) and (d).

\begin{figure}
  \vspace{10pt}
  \centering
  \begin{minipage}{0.48\linewidth}
    \centering
    \begin{overpic}[width=\linewidth]{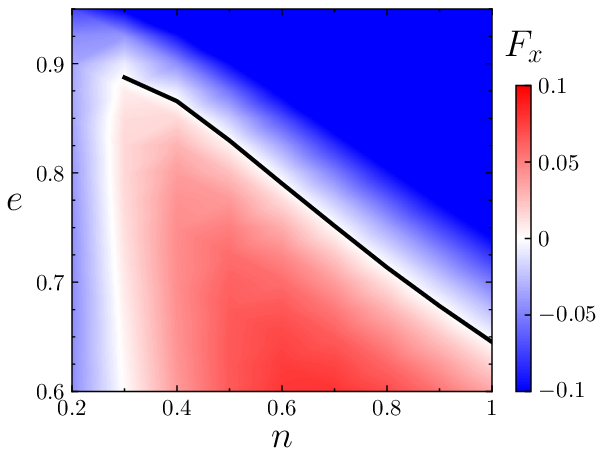}
      \put(5,77){\large (a) $~~\Omega / \omega = 0.02$}
    \end{overpic}
  \end{minipage}
  \hspace{3pt}
  \begin{minipage}{0.48\linewidth}
    \centering
    \begin{overpic}[width=\linewidth]{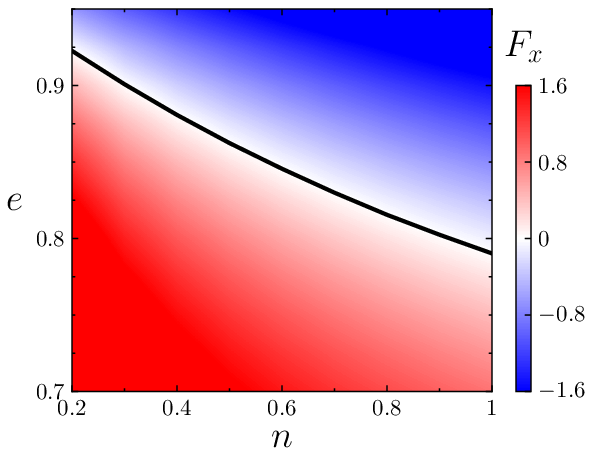}
      \put(5,77){\large (b) $~~\Omega / \omega = 2$}
    \end{overpic}
  \end{minipage}
  \caption{Sign distribution of the lift force $F_x$ and the zero-lift boundary in the $(n,e)$ plane for $\Omega/\omega =$ (a) $0.02$ and (b) $2$. Red and blue indicate $F_x>0$ (toward the narrow gap) and $F_x<0$ (toward the wide gap), respectively. The solid line indicates the zero-lift boundary obtained by linear interpolation of the computed $F_x$ values. }
  \label{fig:fig6}
\end{figure}

To show more clearly how the zero-lift eccentricity varies with $n$, figure~\ref{fig:fig6} presents the sign distribution of the lift force $F_x$ and its zero boundary in the $(n,e)$ plane. 
Red and blue denote $F_x>0$ (toward the narrow gap) and $F_x<0$ (toward the wide gap), respectively, and the solid line indicates the zero-lift boundary obtained by linear interpolation of $F_x$. 
In figure~\ref{fig:fig6}(a), since the zero-lift eccentricity is $e=0$ at $n=0.2$, the solid line is truncated at $n=0.3$. 
In both figure~\ref{fig:fig6}(a) and (b), it can be seen that the zero-lift boundary shifts toward higher $e$ as $n$ decreases. 
It is clearly seen that, at high eccentricity, the lift, which is negative in the Newtonian fluid, can reverse sign and become positive in shear-thinning fluids. 
This suggests that the non-uniform viscosity distribution induced by shear-thinning behaviour has an essential influence on the sign of the lift. 
In the next section, \S~\ref{ssec:mechanism_e}, we first examine, in the Newtonian fluid, the relation between lift reversal and the flow as the eccentricity $e$ varies, 
and then, based on this understanding, in \S~\ref{ssec:mechanism_n}, we quantitatively clarify how the flow-field structures responsible for lift generation are modulated by shear-thinning behaviour, and how this modulation leads to lift reversal.

\subsection{Mechanism of lift reversal induced by eccentricity}
\label{ssec:mechanism_e}

\begin{figure}
  \vspace{15pt}
  \centering
  \begin{minipage}{0.48\linewidth}
    \centering
    \begin{overpic}[width=\linewidth]{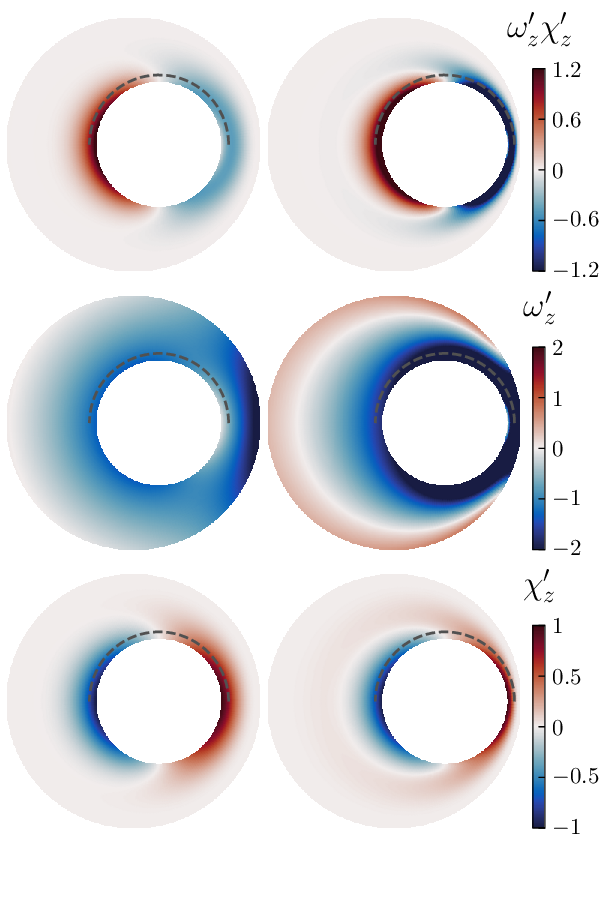}
      \put(0, 103){\large (a)}
      \put(9, 100){\large $e=0.4$}
      \put(39, 100){\large $e=0.8$}
    \end{overpic}
  \end{minipage}
  \hspace{3pt}
  \begin{minipage}{0.48\linewidth}
    \centering
    \begin{overpic}[width=\linewidth]{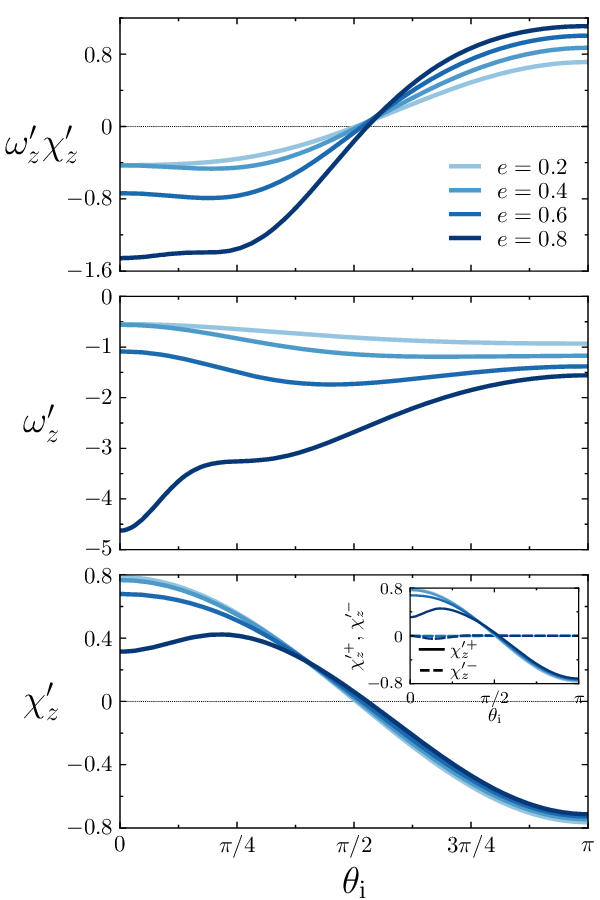}
      \put(0, 103){\large (b)}
      \put(47,41){\includegraphics[width=0.22\linewidth]{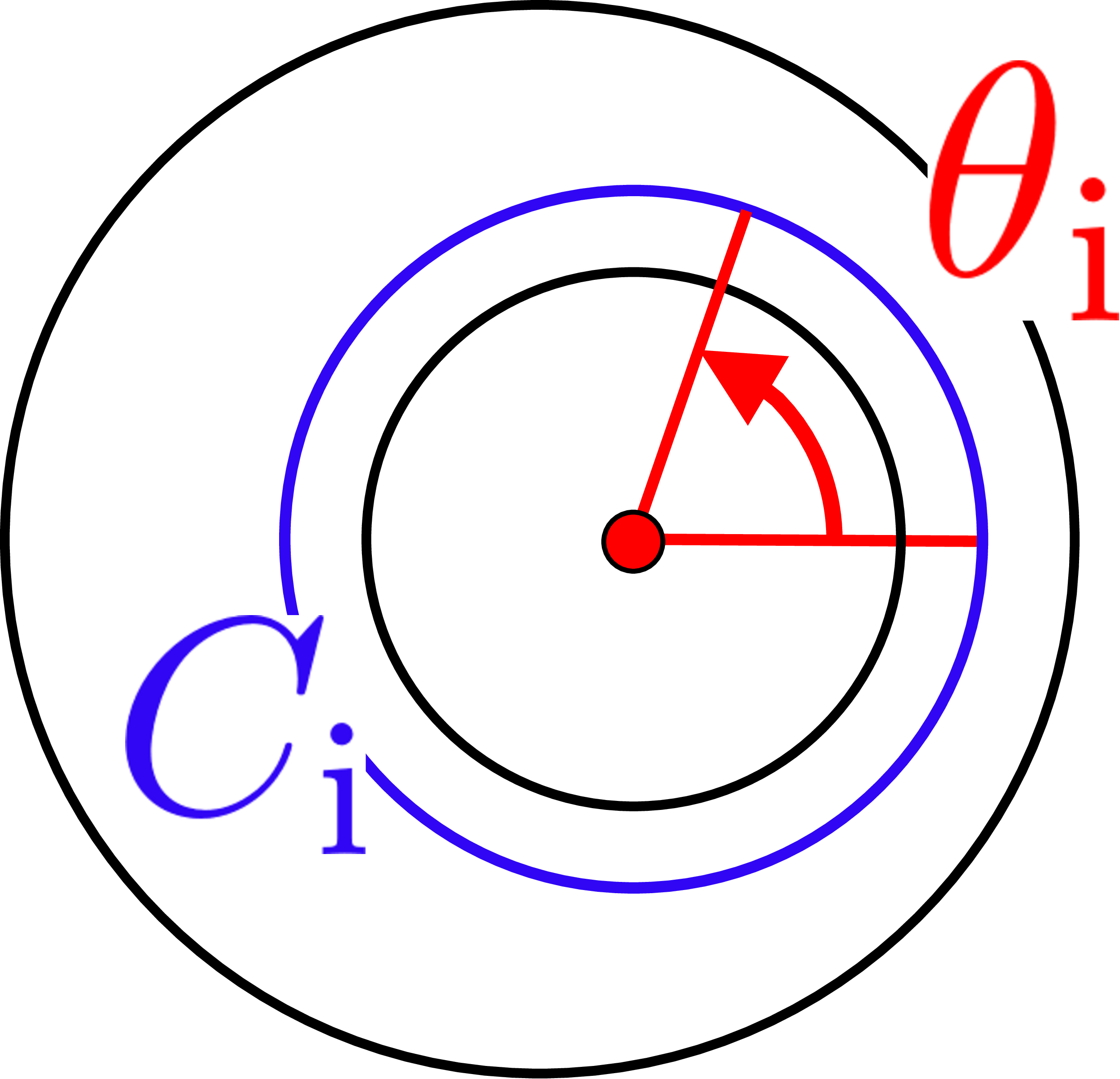}}
    \end{overpic}
  \end{minipage}
  \caption{Visualisation and quantification of the integrand of the vortex-force contribution, $\hat{\boldsymbol{u}}\cdot\boldsymbol{l}'=\omega'_z\chi'_z$, at $\Omega/\omega=0.02$ and $n=1$. Panel (a) shows the spatial distributions of $\omega'_z\chi'_z$, $\omega'_z$, and $\chi'_z=(\hat{\boldsymbol{u}}\times\boldsymbol{u}')\cdot\boldsymbol{e}_z$ for $e=0.4$ and $0.8$. Panel (b) shows the distributions of these quantities as functions of $\theta_\mathrm{i}$, evaluated for each eccentricity $e$ along the portion of the sampling curve $C_\mathrm{i}$ indicated by the grey dashed line in panel (a). Here, $C_\mathrm{i}$ is defined as a circle concentric with the inner cylinder, with radius $R_\mathrm{i}+c(R_\mathrm{o}-R_\mathrm{i})$; in the present figure, $c=0.3$. The angle $\theta_\mathrm{i}$ denotes the azimuthal angle on $C_\mathrm{i}$. The inset for $\chi'_z$ shows the decomposition $\chi'_z=\chi^{\prime+}_z+\chi^{\prime-}_z$, where $\chi^{\prime+}_z=\hat{u}_n u'_t$ and $\chi^{\prime-}_z=-\hat{u}_t u'_n$. Here, $(u'_n,u'_t)$ and $(\hat{u}_n,\hat{u}_t)$ are the normal and tangential components of $\boldsymbol{u}'$ and $\hat{\boldsymbol{u}}$, respectively, on $C_\mathrm{i}$.}
  \label{fig:fig7}
\end{figure}

In this section, we investigate the physical mechanism by which the sign of the lift force $F_x$ reverses with increasing eccentricity $e$ in the Newtonian fluid.
We consider the two cases $\Omega/\omega=0.02$ and $2$, for which $F_x$ varies non-monotonically with eccentricity and exhibits sign reversal, as shown in figure~\ref{fig:fig5}.
Using the volume-integral expression for the lift,~\eqref{eq:frc_v}, we interpret this lift reversal in terms of changes in the flow-field structure.
As shown in~\eqref{eq:frc_v_n1}, the viscous stress contribution $F_x^{(\tau)}$ in~\eqref{eq:frc_v} vanishes for the Newtonian fluid because of geometric symmetry.
We therefore focus in this section on the vortex-force contribution $F_x^{(l)}$.

Firstly, We describe the results for the rotation-dominated condition, $\Omega/\omega=0.02$.
To identify the regions where the lift is generated, we focus on the integrand of the vortex-force contribution $F_x^{(l)}$, namely $\hat{\bm{u}}\cdot\bm{l}'$, that is, the distribution of the local vortex-force contribution.
Figure~\ref{fig:fig7}(a) shows the spatial distribution of $\hat{\bm{u}}\cdot\bm{l}'=\omega'_z\chi'_z$ at $e=0.4$ and $e=0.8$, together with its constituent quantities, the relative vorticity $\omega'_z$ and the weighted relative velocity $\chi'_z=(\hat{\bm{u}}\times\bm{u}')\cdot\bm{e}_z$.
From the upper row of figure~\ref{fig:fig7}(a), which shows $\omega'_z\chi'_z$, it can be seen that, at $e=0.4$, the strong positive contribution distributed in the wide gap region dominates over the negative contribution distributed in the narrow gap region.
By contrast, at $e=0.8$, a pronounced negative contribution appears in the narrow gap region, and this contribution is stronger than that at $e=0.4$.
In other words, at low eccentricity, the positive local vortex-force contribution in the wide gap region keeps $F_x^{(l)}$ positive, whereas, at high eccentricity, the negative local vortex-force contribution in the narrow gap region causes the sign of $F_x^{(l)}$ to become negative.

To clarify the cause of the increase in the negative local vortex-force contribution in the narrow gap region,
we introduce a representative sampling curve $C_\mathrm{i}$ near the inner cylinder and quantitatively examine how the distributions of the relevant quantities along a part of this curve vary with eccentricity.
Here, $C_\mathrm{i}$ is defined as a circle concentric with the inner cylinder, with radius $R_\mathrm{i}+c(R_\mathrm{o}-R_\mathrm{i})$ (with $c=0.1$ in the present figure), and $\theta_\mathrm{i}$ denotes the azimuthal angle on $C_\mathrm{i}$ (a schematic is shown in the middle row of figure~\ref{fig:fig7}(b)).
Figure~\ref{fig:fig7}(b) shows $\omega'_z\chi'_z$, $\omega'_z$, and $\chi'_z$, evaluated over the interval $0\leq\theta_\mathrm{i}\leq\pi$ on $C_\mathrm{i}$ (corresponding to the grey dashed line in figure~\ref{fig:fig7}(a)), as functions of $\theta_\mathrm{i}$.
Focusing on the upper row of figure~\ref{fig:fig7}(b), which shows $\omega'_z\chi'_z$, it can be seen, consistently with the trend suggested by the visualisation in figure~\ref{fig:fig7}(a), that, as $e$ increases, the increase in the negative local vortex-force contribution in the narrow gap region $(0\leq \theta_\mathrm{i}\leq \pi/2)$ exceeds the increase in the positive contribution in the wide gap region $(\pi/2\leq \theta_\mathrm{i}\leq \pi)$.
To identify the origin of this increase, we next examine the distributions of $\omega'_z$ in the middle row and $\chi'_z$ in the lower row of figure~\ref{fig:fig7}(b).
Over the range $0\leq \theta_\mathrm{i}\leq \pi/2$, corresponding to the narrow gap region, $\omega'_z$ increases markedly in the negative direction as $e$ increases, whereas $\chi'_z$ instead decreases slightly.
Therefore, the increase in the negative vortex-force contribution in the narrow gap region can be attributed mainly to the increase in the relative vorticity $\omega'_z$.
This change in $\omega'_z$ corresponds to the fact that, as the eccentricity increases, the outer wall approaches the inner cylinder and the shear induced by the inner-cylinder rotation is amplified within the narrow gap.
For reference, the inset in the lower row of figure~\ref{fig:fig7}(b) shows the decomposition $\chi'_z=\chi_z^{\prime+}+\chi_z^{\prime-}$.
Here, $\chi_z^{\prime+}=\hat{u}_n u'_t$ and $\chi_z^{\prime-}=-\hat{u}_t u'_n$, while $\bm{u}'=(u'_n,u'_t)$ and $\hat{\bm{u}}=(\hat{u}_n,\hat{u}_t)$ denote the normal and tangential components on $C_\mathrm{i}$, respectively.
This decomposition shows that $\chi'_z$ is governed mainly by $\chi_z^{\prime+}$, namely, by the term containing the tangential component of the relative velocity, $u'_t$, although its variation with increasing $e$ is less pronounced than that of $\omega'_z$.
As the gap narrows, the negative relative vorticity therein becomes stronger and strengthens the negative local vortex-force contribution, resulting in the eccentricity-induced lift reversal.

\begin{figure}
  \vspace{15pt}
  \centering
  \begin{minipage}{0.48\linewidth}
    \centering
    \begin{overpic}[width=\linewidth]{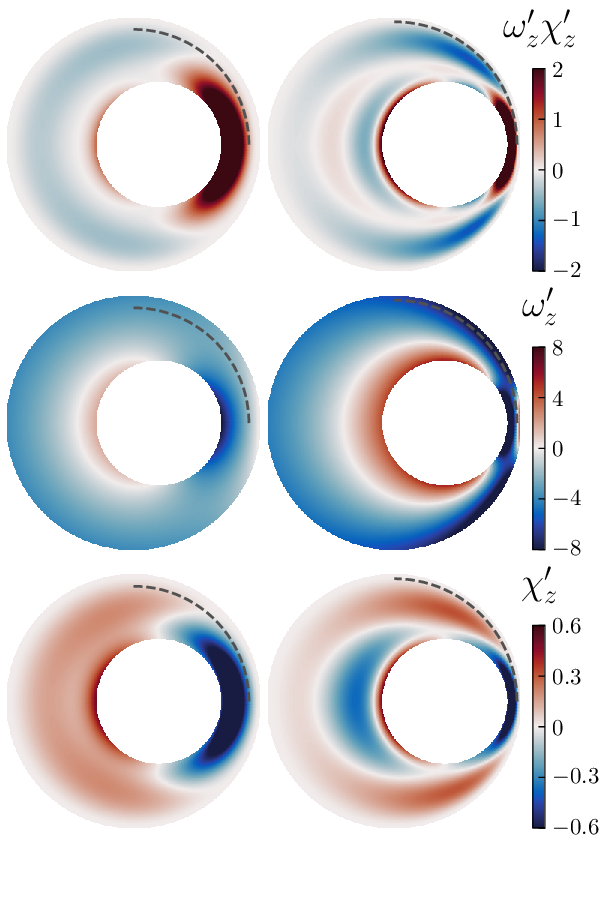}
      \put(0, 103){\large (a)}
      \put(9, 100){\large $e=0.4$}
      \put(39, 100){\large $e=0.8$}
    \end{overpic}
  \end{minipage}
  \hspace{3pt}
  \begin{minipage}{0.48\linewidth}
    \centering
    \begin{overpic}[width=\linewidth]{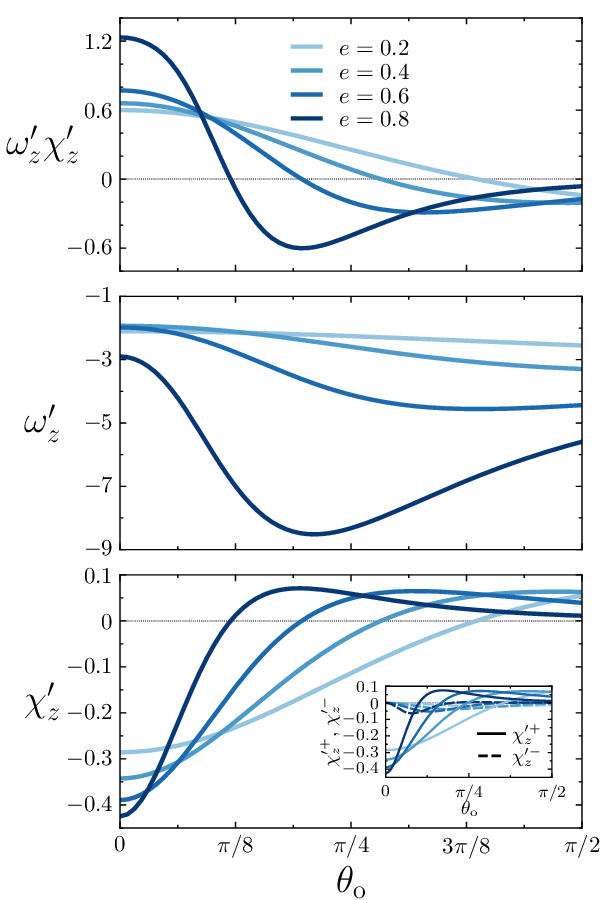}
      \put(0, 103){\large (b)}
      \put(48,83){\includegraphics[width=0.2\linewidth]{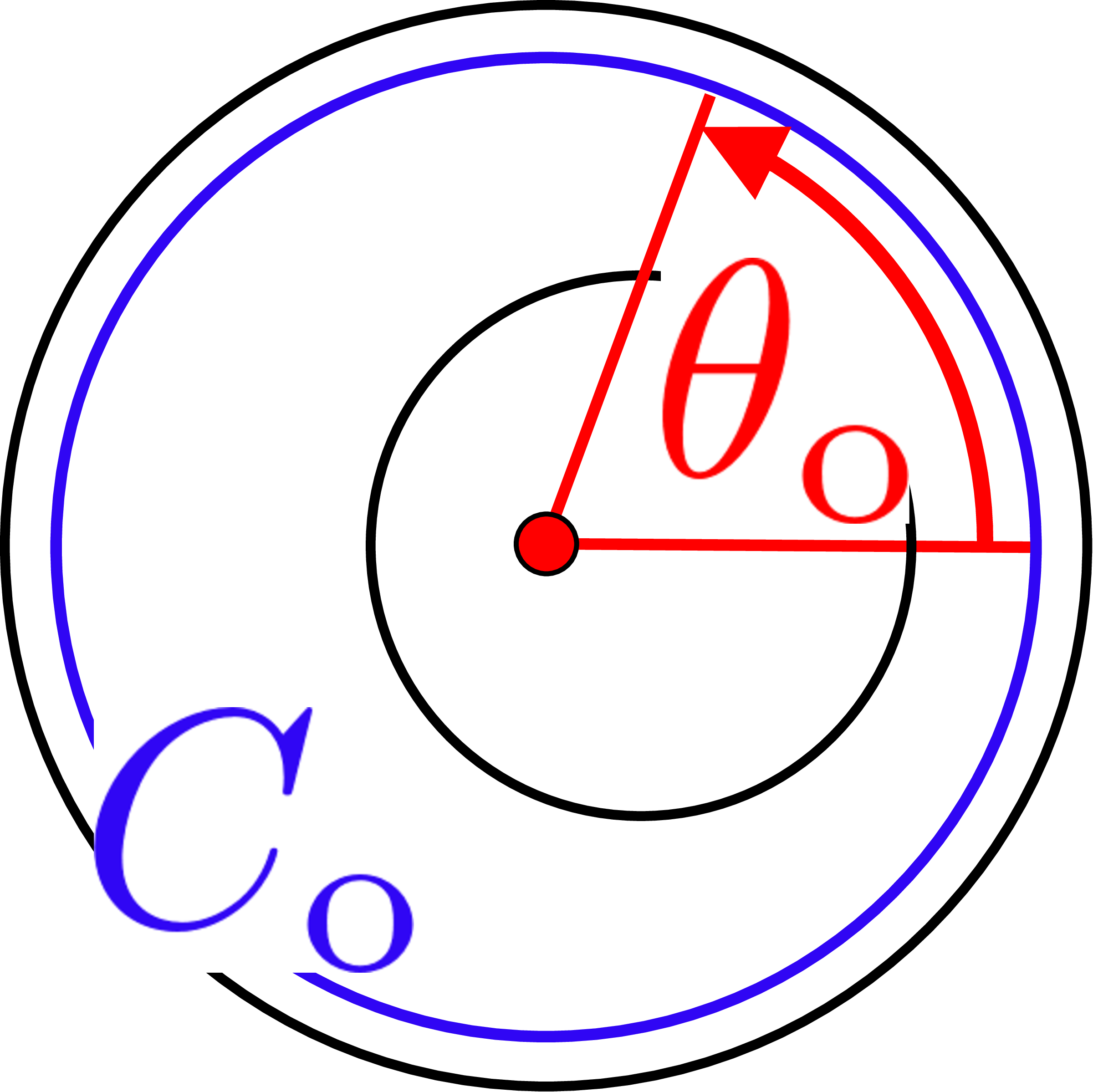}}
    \end{overpic}
  \end{minipage}
  \caption{Visualisation and quantification of the integrand of the vortex-force contribution, $\hat{\boldsymbol{u}}\cdot\boldsymbol{l}'=\omega'_z\chi'_z$, at $\Omega/\omega=2$ and $n=1$. Panel (a) shows the spatial distributions of $\omega'_z\chi'_z$, $\omega'_z$, and $\chi'_z=(\hat{\boldsymbol{u}}\times\boldsymbol{u}')\cdot\boldsymbol{e}_z$ for $e=0.4$ and $0.8$. Panel (b) shows the distributions of these quantities as functions of $\theta_\mathrm{o}$, evaluated for each eccentricity $e$ along the portion of the sampling curve $C_\mathrm{o}$ indicated by the grey dashed line in panel (a). Here, $C_\mathrm{o}$ is defined as a circle concentric with the outer cylinder, with radius $R_\mathrm{o}-c(R_\mathrm{o}-R_\mathrm{i}-\varepsilon)$; in the present figure, $c=0.1$. The angle $\theta_\mathrm{o}$ denotes the azimuthal angle on $C_\mathrm{o}$. The inset for $\chi'_z$ shows the decomposition $\chi'_z=\chi^{\prime+}_z+\chi^{\prime-}_z$, where $\chi^{\prime+}_z=\hat{u}_n u'_t$ and $\chi^{\prime-}_z=-\hat{u}_t u'_n$. Here, $(u'_n,u'_t)$ and $(\hat{u}_n,\hat{u}_t)$ are the normal and tangential components of $\boldsymbol{u}'$ and $\hat{\boldsymbol{u}}$, respectively, on $C_\mathrm{o}$.}
  \label{fig:fig8}
\end{figure}

Secondly, we describe the results for the orbital-dominated condition, $\Omega/\omega=2$.
As in the previous figure, figure~\ref{fig:fig8}(a) shows the spatial distributions of the integrand of the vortex-force contribution, $\hat{\bm{u}}\cdot\bm{l}'=\omega'_z\chi'_z$, together with $\omega'_z$ and $\chi'_z=(\hat{\bm{u}}\times\bm{u}')\cdot\bm{e}_z$.
From the upper row of figure~\ref{fig:fig8}(a), which shows $\omega'_z\chi'_z$, it can be seen that, at $e=0.4$, a strong positive contribution appears in the narrow gap region, whereas, at $e=0.8$, a negative contribution appears in the upper and lower parts of the narrow gap region, particularly near the outer cylinder.
The region where this negative local vortex-force contribution appears is different from that in the rotation-dominated condition, suggesting that, under the orbital-dominated condition, the growth of the negative local vortex-force contribution in this region is the main factor that reverses $F_x^{(l)}$ to negative.

We introduce a representative sampling curve $C_\mathrm{o}$ near the outer cylinder.
Here, $C_\mathrm{o}$ is defined as a circle concentric with the outer cylinder, with radius $R_\mathrm{o}-c(R_\mathrm{o}-R_\mathrm{i}-\varepsilon)$, with $c=0.3$ in this section, and $\theta_\mathrm{o}$ denotes the azimuthal angle on $C_\mathrm{o}$ (a schematic is shown in the middle row of figure~\ref{fig:fig8}(b)).
Figure~\ref{fig:fig8}(b) shows $\omega'_z\chi'_z$, $\omega'_z$, and $\chi'_z$, evaluated over the interval $0\leq\theta_\mathrm{o}\leq\pi/2$ on $C_\mathrm{o}$ (corresponding to the grey dashed line in figure~\ref{fig:fig8}(a)), as functions of $\theta_\mathrm{o}$.
Focusing on the upper row of figure~\ref{fig:fig8}(b), which shows $\omega'_z\chi'_z$, it can be seen, consistently with the trend suggested by the visualisation in figure~\ref{fig:fig8}(a), that, as $e$ increases, the negative local vortex-force contribution increases markedly over $\pi/8\leq\theta_\mathrm{o}\leq\pi/2$, corresponding to the upper and lower parts of the narrow gap region.
To identify the origin of this increase, we next examine the distributions of $\omega'_z$ in the middle row and $\chi'_z$ in the lower row of figure~\ref{fig:fig8}(b).
Over the range $\pi/8\leq\theta_\mathrm{o}\leq\pi/2$, it can be seen that, as $e$ increases, $\omega'_z$ becomes more strongly negative, while $\chi'_z$ changes sign from negative to positive.
The increase of $\omega'_z$ in the negative direction corresponds to the narrowing of the minimum gap with increasing eccentricity, which strengthens the shear in the narrow gap region.
The sign change of $\chi'_z$, on the other hand, arises from the change in the distribution of the tangential relative velocity $u'_t$ near the outer cylinder in the narrow gap region.
Indeed, the inset in the lower row of figure~\ref{fig:fig8}(b) shows that the contribution of $\chi^{\prime+}_z$, which is associated with $u'_t$, plays the dominant role in determining the dependence of $\chi'_z$ on $e$.
The reason why this $u'_t$-related contribution changes can be understood as follows.
When the eccentricity is small, as shown in figure~\ref{fig:fig3}(b), a pressure-driven flow opposite to the shear flow induced by the motion of the inner cylinder, namely a reverse flow, forms near the outer cylinder in the narrow gap region.
By contrast, at high eccentricity, the region in which this pressure-driven flow develops significantly becomes narrower, and the reverse flow near the outer cylinder is therefore suppressed.
As a result, the tangential relative velocity develops more clearly near the outer cylinder in the narrow gap region, and this effect appears as a positive increase in $\chi'_z$.
Near the outer cylinder in the narrow gap region, increasing $e$ causes the concurrent enhancement of negative relative vorticity and tangential velocity. Such a combined effect strengthens the negative local vortex-force contribution, resulting in the eccentricity-induced lift reversal.

\subsection{Mechanism of lift reversal induced by shear-thinning behaviour}
\label{ssec:mechanism_n}

In this section, we investigate the physical mechanism by which the sign of the lift force $F_x$ reverses from negative to positive, under high-eccentricity conditions, as the power-law index $n$ decreases, that is, as shear-thinning behaviour becomes stronger, as shown in figure~\ref{fig:fig6}.
We consider the two cases, $\Omega/\omega=0.02$ and $2$, in which such $n$-dependent lift reversal appears clearly.
In the previous section, for the negative lift generated in the Newtonian fluid at high eccentricity, we showed that strong development of relative vorticity in the narrow gap region is important under the rotation-dominated condition, whereas under the orbital-dominated condition, changes in the tangential relative velocity also play an essential role in addition to this effect.
In this section, based on the proposed lift-diagnostic framework, we clarify how the flow-field features identified in the previous section as being important for the generation of negative lift are modulated by the non-uniform viscosity distribution arising from shear-thinning behaviour, and how this modulation leads to the sign reversal of the lift.

\begin{figure}
  \vspace{10pt}
  \centering
  \begin{minipage}[t]{0.51\textwidth}
    \vspace{0pt}
    \centering
    \begin{overpic}[width=\linewidth]{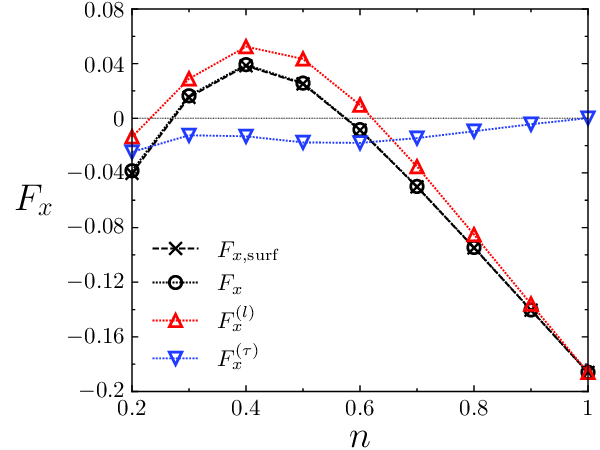}
      \put(15,78){\large (a) $~~\Omega / \omega = 0.02$}
    \end{overpic}
  \end{minipage}%
  \hspace{-15pt}%
  \begin{minipage}[t]{0.51\textwidth}
    \vspace{0pt}
    \centering
    \begin{overpic}[width=\linewidth]{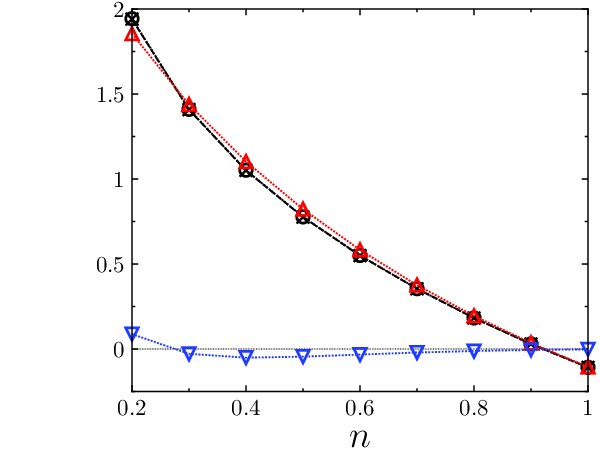}
      \put(15,78){\large (b) $~~\Omega / \omega = 2$}
    \end{overpic}
  \end{minipage}
  \caption{Decomposition of the lift force $F_x$ in a power-law fluid at $e = 0.8$ for two rotation-rate ratios, based on equation \eqref{eq:frc_v}. Panels (a) and (b) show the dependence of the lift and its decomposed components on the power-law index $n$ for (a) $\Omega/\omega = 0.02$ and (b) $\Omega/\omega = 2$, respectively. Here, $F_x^{(l)}$ and $F_x^{(\tau)}$ denote the vortex-force and shear-stress contributions to the lift, respectively. $F_{x,\mathrm{surf}}$ represents the lift evaluated from the surface integral \eqref{eq:frc_s}.}
  \label{fig:fig9}
\end{figure}

We first decompose the lift force $F_x$ based on~\eqref{eq:frc_v} in order to identify the dominant contribution to the lift in shear-thinning fluids.
Figure~\ref{fig:fig9} shows the lift force $F_x$ at the rotation-rate ratios $\Omega/\omega=$ (a) $0.02$ and (b) $2$, with eccentricity $e=0.8$, decomposed into the vortex-force contribution $F_x^{(l)}$ and the viscous stress contribution $F_x^{(\tau)}$, as functions of the power-law index $n$.
For comparison, the lift $F_{x,\mathrm{surf}}$ evaluated from the surface integral~\eqref{eq:frc_s} is also shown.
Under both conditions, $F_{x,\mathrm{surf}}$ and $F_x$ evaluated from~\eqref{eq:frc_v} agree well over the entire range of $n$, confirming that the volume-integral representation based on~\eqref{eq:frc_v} provides an accurate evaluation of the lift.
We next focus on the $n$-dependence of $F_x$.
For $\Omega/\omega=0.02$ in panel (a), $F_x$ first increases as $n$ decreases and becomes positive over an intermediate range of $n$, whereas it becomes negative again in the strongly shear-thinning regime $(n<0.3)$.
For $\Omega/\omega=2$ in panel (b), $F_x$ increases almost monotonically as $n$ decreases, and changes sign from negative to positive at $n\leq0.9$.
Focusing on the decomposition of $F_x$ into its two contributions, we find that, under both conditions, $F_x^{(\tau)}$ owing to the viscosity variation, remains much smaller than $F_x$ over the entire range and shows only weak dependence on $n$.
By contrast, $F_x^{(l)}$ has nearly the same magnitude and $n$-dependence as $F_x$, indicating that the lift variation induced by shear-thinning behaviour is governed mainly by the vortex-force contribution.
Accordingly, in what follows, we focus on the integrand of the vortex-force contribution.
For $\Omega/\omega=0.02$, $F_x$ becomes negative again in the strongly shear-thinning regime $(n<0.3)$; however, because this secondary reversal is not the main focus of the present section, the following mechanism analysis is restricted to $n\geq0.4$, which corresponds to the range of the primary sign reversal.

\begin{figure}
  \vspace{15pt}
  \centering
  \begin{minipage}{0.48\linewidth}
    \centering
    \begin{overpic}[width=\linewidth]{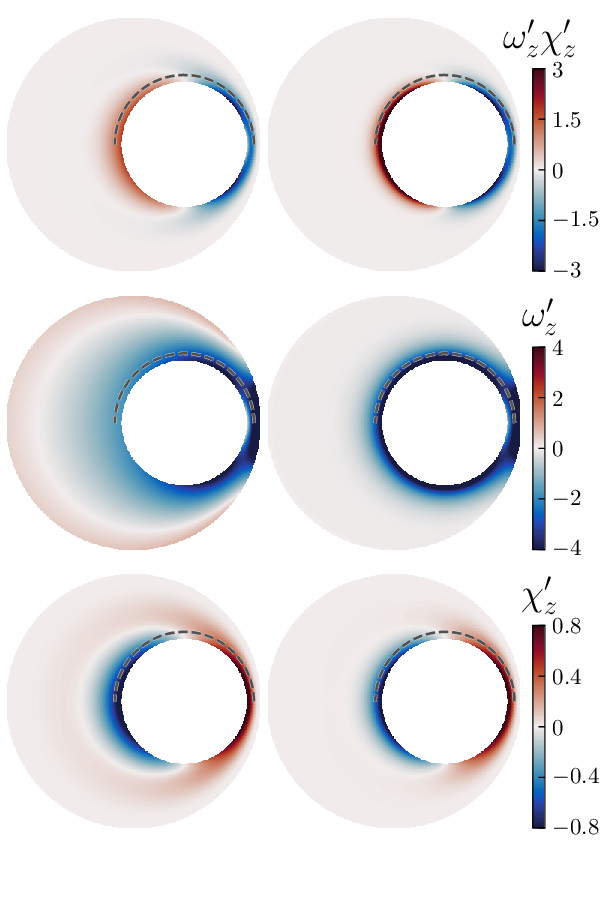}
      \put(0, 103){\large (a)}
      \put(11, 100){\large $n=1$}
      \put(39, 100){\large $n=0.4$}
    \end{overpic}
  \end{minipage}
  \hspace{3pt}
  \begin{minipage}{0.48\linewidth}
    \centering
    \begin{overpic}[width=\linewidth]{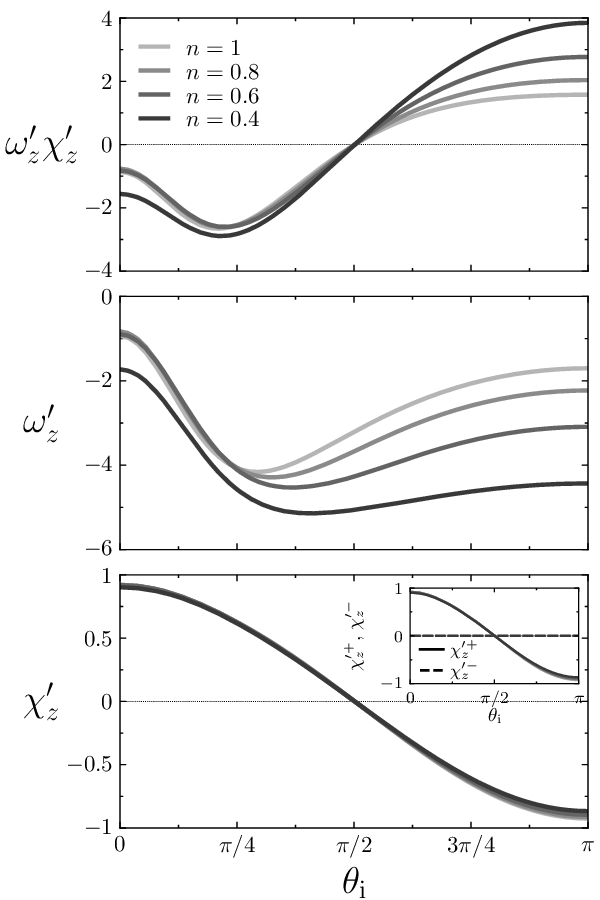}
      \put(0, 103){\large (b)}
      \put(48,72){\colorbox{white}{\includegraphics[width=0.22\linewidth]{ci.eps}}}
    \end{overpic}
  \end{minipage}
  \caption{Visualisation and quantification of the integrand of the vortex-force contribution, $\hat{\boldsymbol{u}}\cdot\boldsymbol{l}'=\omega'_z\chi'_z$, at $\Omega/\omega=0.02$ and $e=0.8$.
Panel (a) shows the spatial distributions of $\omega'_z\chi'_z$, $\omega'_z$, and $\chi'_z=(\hat{\boldsymbol{u}}\times\boldsymbol{u}')\cdot\boldsymbol{e}_z$ for $n=1.0$ and $0.4$.
Panel (b) shows these quantities as functions of $\theta_\mathrm{i}$ for each $n$, evaluated along the portion of the sampling curve $C_\mathrm{i}$ indicated by the grey dashed line in panel (a).
Here, $C_\mathrm{i}$ is defined as a circle concentric with the inner cylinder, with radius $R_\mathrm{i}+c(R_\mathrm{o}-R_\mathrm{i})$; in the present figure, $c=0.5$, and $\theta_\mathrm{i}$ denotes the azimuthal angle on $C_\mathrm{i}$.
The inset in the lower row of panel (b) shows the decomposition $\chi'_z=\chi_z^{\prime +}+\chi_z^{\prime -}$, where $\chi_z^{\prime +}=\hat{u}_n u'_t$ and $\chi_z^{\prime -}=-\hat{u}_t u'_n$.
Here, $(u'_n,u'_t)$ and $(\hat{u}_n,\hat{u}_t)$ are the normal and tangential components of $\boldsymbol{u}'$ and $\hat{\boldsymbol{u}}$, respectively, on $C_\mathrm{i}$.}
  \label{fig:fig10}
\end{figure}

Firstly, We investigate the mechanism of lift reversal induced by increasing shear-thinning behaviour under the rotation-dominated condition, $\Omega/\omega=0.02$, at the high eccentricity $e=0.8$.
As in the previous section, figure~\ref{fig:fig10}(a) shows the spatial distribution of the integrand of the vortex-force contribution, $\hat{\bm{u}}\cdot\bm{l}'=\omega'_z\chi'_z$, together with its constituent quantities, the relative vorticity $\omega'_z$ and the weighted relative velocity $\chi'_z=(\hat{\bm{u}}\times\bm{u}')\cdot\bm{e}_z$.
Focusing on the upper row of figure~\ref{fig:fig10}(a), which shows $\omega'_z\chi'_z$, it can be seen that, at $n=1$, positive contributions are distributed in the wide gap region and negative contributions in the narrow gap region, with the latter being relatively dominant.
By contrast, at $n=0.4$, the positive contribution localised near the inner cylinder in the wide gap region increases markedly.
This suggests that the increase in the positive local vortex-force contribution in this region is the main factor that reverses $F_x^{(l)}$ from negative to positive.

We shall discuss the cause of the increase in the positive local vortex-force contribution near the inner cylinder in the wide gap region using $C_\mathrm{i}$, in the same manner as in \S~\ref{ssec:mechanism_e} but with $c=0.5$.
The angle $\theta_\mathrm{i}$ denotes the azimuthal angle on $C_\mathrm{i}$ (a schematic is shown in the middle row of figure~\ref{fig:fig10}(b)).
Figure~\ref{fig:fig10}(b) shows $\omega'_z\chi'_z$, $\omega'_z$, and $\chi'_z$, evaluated over the interval $0\leq\theta_\mathrm{i}\leq\pi$ on $C_\mathrm{i}$ (corresponding to the grey dashed line in figure~\ref{fig:fig10}(a)), as functions of $\theta_\mathrm{i}$.
Focusing on the upper row of figure~\ref{fig:fig10}(b), which shows $\omega'_z\chi'_z$, it is found that, as $n$ decreases, both the negative contribution in the narrow gap region $(0\leq\theta_\mathrm{i}\leq\pi/2)$ and the positive contribution in the wide gap region $(\pi/2\leq\theta_\mathrm{i}\leq\pi)$ increase, but that the latter increase is more pronounced.
To clarify the origin of this increase, we next examine the distributions of $\omega'_z$ in the middle row and $\chi'_z$ in the lower row of figure~\ref{fig:fig10}(b).
Over the range $\pi/2\leq\theta_\mathrm{i}\leq\pi$, corresponding to the wide gap region, it can be seen that, as $n$ decreases, $\omega'_z$ near the inner cylinder becomes more strongly negative, whereas the change in $\chi'_z$ is relatively small.
Therefore, the increase in the positive local vortex-force contribution in this region can be interpreted as arising mainly from the increase in the magnitude of the relative vorticity $\omega'_z$.
Indeed, the spatial distributions in figure~\ref{fig:fig10}(a) also confirm that, at $n=0.4$, negative vorticity develops more strongly near the inner cylinder in the wide gap region.
This amplification of the relative vorticity is attributed to the viscosity reduction near the inner cylinder, where the shear flow forms, which causes the velocity field to vary more sharply.
Stronger shear-thinning behaviour intensifies negative relative vorticity near the inner cylinder in the wide gap region, thereby increasing the positive local vortex-force contribution and causing the lift to reverse.

\begin{figure}
  \vspace{15pt}
  \centering
  \begin{minipage}{0.48\linewidth}
    \centering
    \begin{overpic}[width=\linewidth]{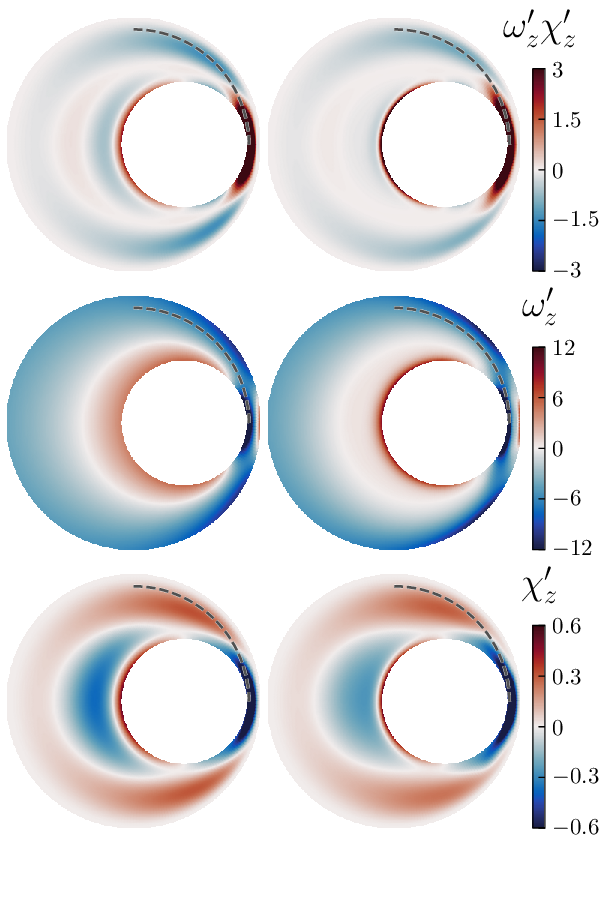}
      \put(0, 103){\large (a)}
      \put(11, 100){\large $n=1$}
      \put(39, 100){\large $n=0.4$}
    \end{overpic}
  \end{minipage}
  \hspace{3pt}
  \begin{minipage}{0.48\linewidth}
    \centering
    \begin{overpic}[width=\linewidth]{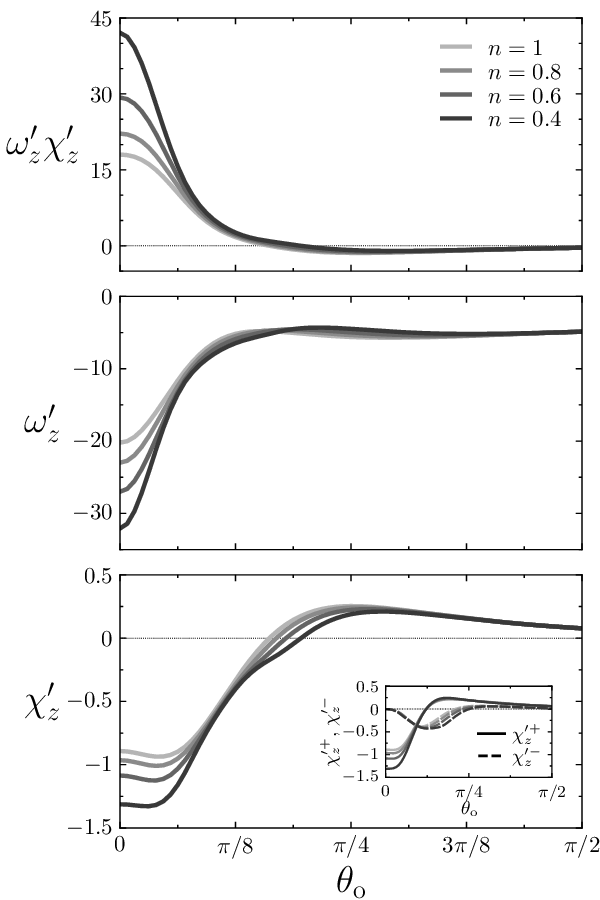}
      \put(0, 103){\large (b)}
      \put(47,43){\includegraphics[width=0.22\linewidth]{co.eps}}
    \end{overpic}
  \end{minipage}
  \caption{Visualisation and quantification of the integrand of the vortex-force contribution, $\hat{\boldsymbol{u}}\cdot\boldsymbol{l}'=\omega'_z\chi'_z$, at $\Omega/\omega=2$ and $e=0.8$.
Panel (a) shows the spatial distributions of $\omega'_z\chi'_z$, $\omega'_z$, and $\chi'_z=(\hat{\boldsymbol{u}}\times\boldsymbol{u}')\cdot\boldsymbol{e}_z$ for $n=1.0$ and $0.4$.
Panel (b) shows these quantities as functions of $\theta_\mathrm{o}$ for each $n$, evaluated along the portion of the sampling curve $C_\mathrm{o}$ indicated by the grey dashed line in panel (a).
Here, $C_\mathrm{o}$ is defined as a circle concentric with the outer cylinder, with radius $R_\mathrm{o}-c(R_\mathrm{o}-R_\mathrm{i}-\varepsilon)$; in the present figure, $c=0.9$, and $\theta_\mathrm{o}$ denotes the azimuthal angle on $C_\mathrm{o}$.
The inset in the lower row of panel (b) shows the decomposition $\chi'_z=\chi_z^{\prime +}+\chi_z^{\prime -}$, where $\chi_z^{\prime +}=\hat{u}_n u'_t$ and $\chi_z^{\prime -}=-\hat{u}_t u'_n$.
Here, $(u'_n,u'_t)$ and $(\hat{u}_n,\hat{u}_t)$ are the normal and tangential components of $\boldsymbol{u}'$ and $\hat{\boldsymbol{u}}$, respectively, on $C_\mathrm{o}$.}
  \label{fig:fig11}
\end{figure}

Secondly, We investigate the mechanism of lift reversal induced by increasing shear-thinning behaviour under the orbital-dominated condition, $\Omega/\omega=2$, at the high eccentricity $e=0.8$.
As in the previous figure, figure~\ref{fig:fig11}(a) shows the spatial distribution of the integrand of the vortex-force contribution, $\hat{\bm{u}}\cdot\bm{l}'=\omega'_z\chi'_z$, together with its constituent quantities, $\omega'_z$ and $\chi'_z=(\hat{\bm{u}}\times\bm{u}')\cdot\bm{e}_z$.
Focusing on the upper row of figure~\ref{fig:fig11}(a), which shows $\omega'_z\chi'_z$, it can be seen that, at $n=1$, a strong positive contribution is localised near the minimum gap, while negative contributions are distributed over a relatively broad region near the outer cylinder in the upper and lower parts of the narrow gap region.
By contrast, at $n=0.4$, these negative contributions in the upper and lower parts become relatively weaker.
As a supplementary remark, although this is not emphasised in the visualisation, the maximum positive contribution in the narrow gap region also increases, by about a factor of two at $n=0.4$ compared with that at $n=1.0$.
These observations suggest that the lift reversal induced by increasing shear-thinning behaviour at $\Omega/\omega=2$ and $e=0.8$ is caused by both the reduction of the negative local vortex-force contribution in the upper and lower parts of the narrow gap region, which is dominant in the Newtonian fluid, and the increase in the positive contribution near the minimum gap.

To confirm this quantitatively, we introduce a sampling curve $C_\mathrm{o}$ near the outer cylinder, in the same manner as in \S~\ref{ssec:mechanism_e} but with $c=0.9$.
The angle $\theta_\mathrm{o}$ denotes the azimuthal angle on $C_\mathrm{o}$ (a schematic is shown in the middle row of figure~\ref{fig:fig11}(b)).
Figure~\ref{fig:fig11}(b) shows $\omega'_z\chi'_z$, $\omega'_z$, and $\chi'_z$, evaluated over the interval $0\leq\theta_\mathrm{o}\leq\pi/2$ on $C_\mathrm{o}$ (corresponding to the grey dashed line in figure~\ref{fig:fig11}(a)), as functions of $\theta_\mathrm{o}$.
Focusing on the upper row of figure~\ref{fig:fig11}(b), which shows $\omega'_z\chi'_z$, it can be seen that, as $n$ decreases, the positive local vortex-force contribution over $0\leq\theta_\mathrm{o}\leq\pi/8$, corresponding to the vicinity of the minimum gap, increases markedly.
By contrast, the change in the negative contribution in the upper and lower parts of the narrow gap region $(\pi/4\leq\theta_\mathrm{o}\leq\pi/2)$ is relatively small compared with the increase in this positive contribution.
Therefore, the lift reversal induced by increasing shear-thinning behaviour at $\Omega/\omega=2$ and $e=0.8$ can be attributed mainly to the increase in the positive local vortex-force contribution near the minimum gap.

To clarify the origin of this increase in the positive local vortex-force contribution, we focus on the distributions of $\omega'_z$ and $\chi'_z$ shown in the middle and lower rows of figure~\ref{fig:fig11}(b).
Over the range $0\leq\theta_\mathrm{o}\leq\pi/8$ near the minimum gap, it can be seen that, as $n$ decreases, $\omega'_z$ increases markedly in the negative direction, and $\chi'_z$ also becomes more negative.
As a result, the positive peak of $\omega'_z\chi'_z$ is strongly amplified as the product of negative $\omega'_z$ and negative $\chi'_z$.
Therefore, the increase in the positive local vortex-force contribution in this region can be interpreted as arising mainly from the increase in the magnitude of the relative vorticity $\omega'_z$, while the change in $\chi'_z$ plays a secondary role that reinforces this effect.
This amplification of the relative vorticity is attributed to the viscosity reduction in the high-shear region near the minimum gap as shear-thinning behaviour becomes stronger, which in turn makes the variation of the velocity field steeper.
At $\Omega/\omega=2$ and $e=0.8$, the vortex-force contribution $\omega^\prime_z \chi^\prime_z$ is markedly positive preferentially in the narrow gap both in the Newtonian ($n=1$) and non-Newtonian ($n=0.4$) cases. Greater shear-thinning intensifies such a preferential effect, outweighing the negative vortex-force contribution in the bulk and thereby inducing the lift reversal.

\section{Conclusion}
\label{sec:con}
In this study, we developed a lift-diagnostic framework based on the generalised reciprocal theorem to interpret the origin of steady inertial lift in wall-bounded flows in relation to the internal structure of the flow field.
We applied this framework to the flow between eccentric rotating cylinders in Newtonian and shear-thinning fluids, and investigated two problems using flow-field data obtained from numerical simulations: lift reversal induced by increasing eccentricity, and lift reversal induced by shear-thinning behaviour.

Within the present framework, the steady lift in this system can be decomposed into a vortex-force contribution arising from finite inertia and a viscous stress contribution arising from the non-uniform viscosity distribution associated with shear-thinning behaviour.
Over the range of parameters considered here, we showed that the variation of lift, particularly its sign reversal, is governed mainly by the vortex-force contribution (figure~\ref{fig:fig9}).
Furthermore, by examining the spatial distribution of the local vortex-force contribution, we identified the regions that play an essential role in lift generation, and by focusing on the relative vorticity and the weighted relative velocity that constitute the integrand, we interpreted the lift-generation mechanism in relation to changes in the flow field.

First, for lift reversal induced by increasing eccentricity in the Newtonian fluid, we visualised the distribution of the local vortex-force contribution based on the proposed framework and interpreted its mechanism in relation to changes in the flow field.
We examined two representative conditions: a rotation-dominated condition ($\Omega/\omega = 0.02$) and an orbital-dominated condition ($\Omega/\omega = 2$).
Under the rotation-dominated condition, we showed that lift reversal is mainly caused by the development of negative relative vorticity in the narrow-gap region as the eccentricity increases, which makes the negative local vortex-force contribution dominant in that region (figure~\ref{fig:fig7}).
Under the orbital-dominated condition, negative relative vorticity likewise develops on the outer-cylinder side of the narrow-gap region, and the weighted relative velocity also increases there.
As a result, the negative local vortex-force contribution becomes dominant, leading to lift reversal (figure~\ref{fig:fig8}).

Next, in the shear-thinning fluid, we showed that the eccentricity at which the lift becomes zero shifts to larger values, and that the sign of the lift can reverse even at the same eccentricity.
To interpret this non-trivial variation, we extended the above analysis for the Newtonian fluid and examined how the relative vorticity and the weighted relative velocity are modified by shear-thinning behaviour.
We found that viscosity reduction in the high-shear region near the inner cylinder markedly enhances the relative vorticity and thereby strengthens the positive local vortex-force contribution, which is the main cause of lift reversal induced by shear-thinning behaviour (figures~\ref{fig:fig10} and \ref{fig:fig11}).

These results demonstrate that the formulation developed in this study provides an effective framework for identifying, directly from the flow field, the regions and dominant contributions essential to the generation of weak steady inertial lift in wall-bounded flows, and for interpreting them in direct relation to the flow structure.
Although the present study considered two-dimensional steady flow around a rigid body in Newtonian and shear-thinning fluids, future extensions to unsteady flows, deformable particles, more complex constitutive laws, and systems involving multiple bodies should further broaden our understanding of lift generation in wall-bounded flows.
Such extensions will not only deepen the systematic understanding of lift in wall-bounded flows, but also contribute to the control of the motion of particles, drops, and bubbles, to the design of lubrication systems and transport processes, and to advanced manipulation in microfluidic devices.

\section*{Acknowledgements.}
\noindent MH was supported by JST SPRING, Grant Number JPMJSP2138. Part of the results was obtained by supercomputer Fugaku at the RIKEN R-CCS through the HPCI System Research Project (Project ID: hp220106) and SQUID at D3 centre, The University of Osaka.
\vspace{-5mm}
\section*{Declaration of interests.}  
\noindent The authors report no conflict of interest.

\appendix

\section{Finite-difference descriptions of the basic equation set
in bipolar coordinates}
\label{sec:appA}
\begin{figure}
  \centering
  \begin{minipage}{0.45\linewidth}
    \centering
    \begin{overpic}[width=\linewidth]{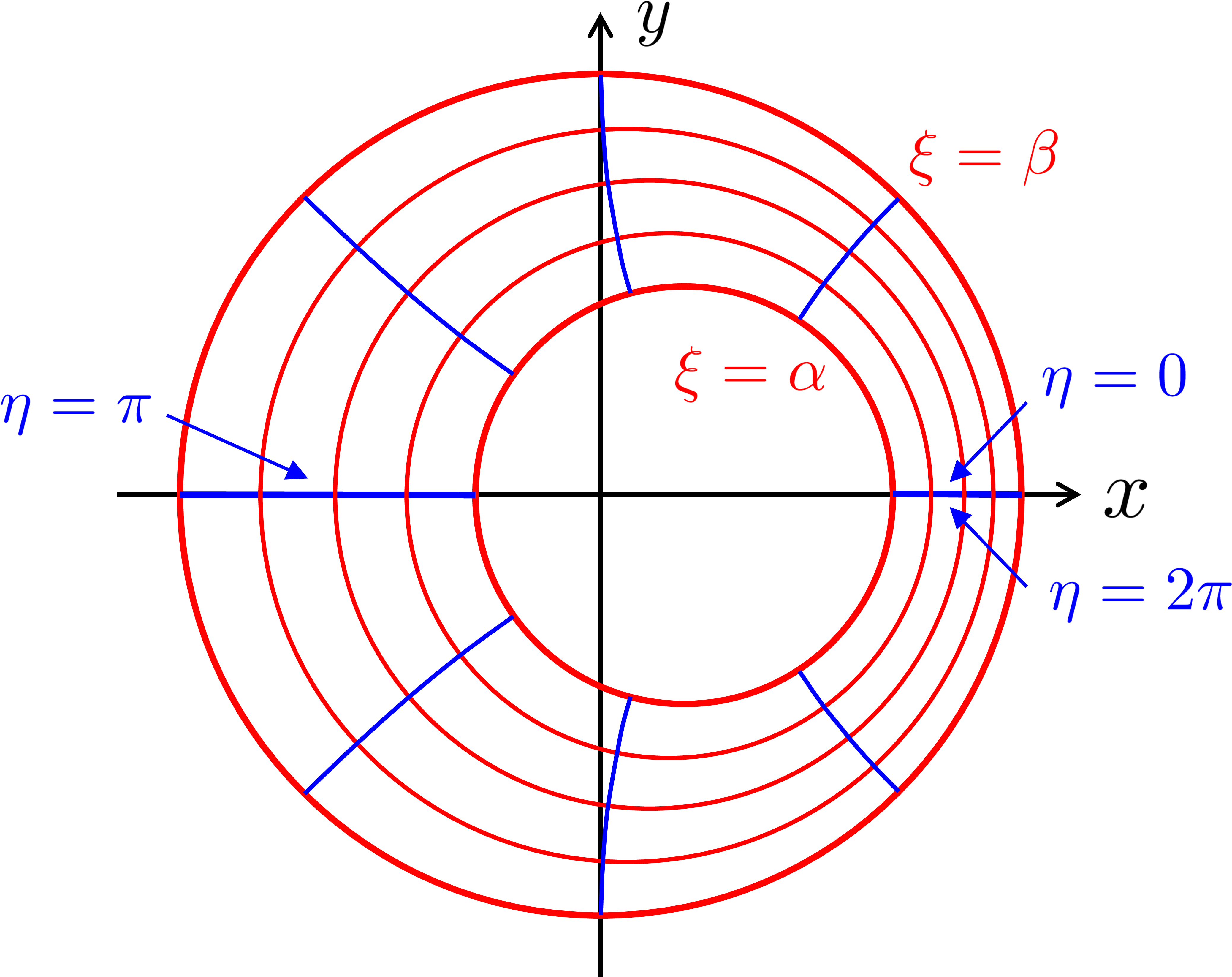}
      \put(0,74){\large (a)}
    \end{overpic}
  \end{minipage}
  \hspace{3pt}
  \begin{minipage}{0.45\linewidth}
    \centering
    \begin{overpic}[width=0.84\linewidth]{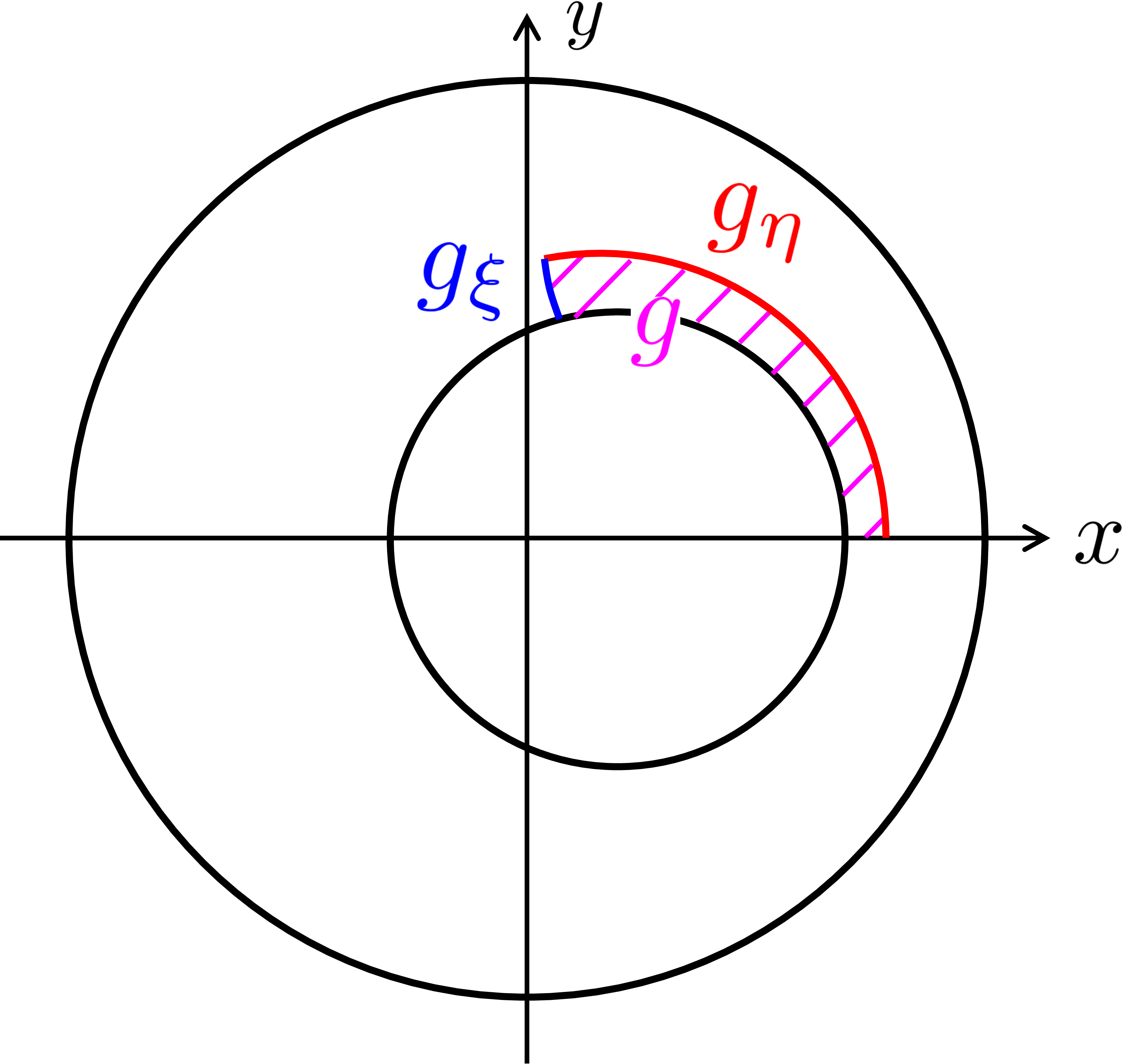}
      \put(0,88){\large (b)}
    \end{overpic}
  \end{minipage}
  \caption{(a) Schematic of the bipolar coordinate system $(\xi,\eta)$. The surfaces of the inner and outer cylinders correspond to $\xi=\alpha$ and $\xi=\beta$, respectively.
(b) Schematic of the curve lengths and area. The quantities $g_{\xi}$ and $g_{\eta}$ denote the curve lengths in the $\xi$ and $\eta$ directions, respectively, and $g$ denotes the area enclosed by them.}
  \label{fig:app1}
\end{figure}

We formulate the governing equations in the bipolar coordinate system $(\xi,\eta)$ shown in figure~\ref{fig:app1}(a)$~$(see \citealp[Appendix A–19]{happel1973low}).
The Cartesian coordinates $(x,y)$ are expressed through the conformal mapping as
\begin{equation}
  x=-\frac{k \sinh (\alpha+\beta-\xi)}{\mathcal{D}}+k \operatorname{coth} \alpha, \quad y=\frac{k \sin \eta}{\mathcal{D}}.
\end{equation}
Here, $\mathcal{D}=\cosh(\alpha+\beta-\xi)+\cos\eta$, with $\xi\in[\alpha,\beta]$ and $\eta\in[0,2\pi]$.
The constants $k$, $\alpha$, and $\beta$ are determined from
\begin{equation}
\begin{gathered}
R_{\mathrm{o}}=\frac{k}{\sinh \alpha}, 
\quad
R_{\mathrm{i}}=\frac{k}{\sinh \beta}, 
\quad
\varepsilon=k\left(\operatorname{coth}\alpha-\operatorname{coth}\beta\right), \\
\cosh \alpha=\frac{R_{\mathrm{o}}^{2}-R_{\mathrm{i}}^{2}+\varepsilon^{2}}{2R_{\mathrm{o}}\varepsilon}, 
\quad
\cosh \beta=\frac{R_{\mathrm{o}}^{2}-R_{\mathrm{i}}^{2}-\varepsilon^{2}}{2R_{\mathrm{i}}\varepsilon},
\end{gathered}
\end{equation}
in terms of the outer and inner cylinder radii $R_{\mathrm{o}}$, $R_{\mathrm{i}}$, and the centre-to-centre distance $\varepsilon$.
The surfaces of the inner and outer cylinders correspond to $\xi=\alpha$ and $\xi=\beta$, respectively.
The scale factors are defined, for example, as $h_{\xi}=\{(\partial x/\partial\xi)^{2}+(\partial y/\partial\xi)^{2}\}^{1/2}$
(with an analogous definition for $h_{\eta}$; see \citealp[Appendix B]{batchelor1967introduction}).
Owing to the conformal nature of the mapping, the scale factors satisfy $h_{\xi}=h_{\eta}$, and are given by
\begin{equation}
  h_{\xi}=h_{\eta}=\frac{k}{\mathcal{D}},\qquad
  h=h_{\xi}h_{\eta}=\frac{k^{2}}{\mathcal{D}^{2}}.
  \label{eq:scale_factor}
\end{equation}

We employ a standard staggered-grid arrangement. Pressure is defined at cell centres, whereas the velocity components are defined on the corresponding cell faces.
The governing equations are spatially discretised using a second-order finite-difference scheme.
To ensure discrete conservation of mass and momentum, and to minimise numerical integration errors in the evaluation of forces and volume-integrated quantities, the physical grid widths and cell areas in the bipolar coordinate system are evaluated using their exact expressions. For this purpose, the integrals of the scale factors
\begin{align}
  &g_{\xi}(\xi, \eta) =\int_{\alpha}^{\xi} \frac{k}{\cosh (\alpha+\beta-\bar{\xi})+\cos \eta} \mathrm{d} \bar{\xi}=-\frac{k\left(\Theta-\Theta_{\mathrm{i}}\right)}{\sin \eta}, \\
  &g_{\eta}(\xi, \eta) =\int_{0}^{\eta} \frac{k}{\cosh (\alpha+\beta-\xi)+\cos \bar{\eta}} \mathrm{d} \bar{\eta}=\frac{k \Theta}{\sinh (\alpha+\beta-\xi)}, \\
  &g(\xi, \eta)  = \int_{0}^{\eta} \int_{\alpha}^{\xi}\left(\frac{k}{\cosh (\alpha+\beta-\bar{\xi})+\cos \bar{\eta}}\right)^{2} \mathrm{~d} \bar{\xi} \mathrm{d} \bar{\eta} \notag \\
  &~~=\frac{k^{2}}{2}\Bigl\{
\Bigl(\frac{1}{\sinh^{2}(\alpha+\beta-\xi)}-\frac{1}{\sin^{2}\eta}\Bigr)\Theta
-\Bigl(\frac{1}{\sinh^{2}\beta}-\frac{1}{\sin^{2}\eta}\Bigr)\Theta_{\mathrm{i}}
+\frac{\mathcal{C}-2}{\mathcal{S}}-\frac{\mathcal{C}_{\mathrm{i}}-2}{\mathcal{S}_{\mathrm{i}}}
\Bigr\},
  \label{eq:length_and_area}
\end{align}
are introduced, as illustrated in figure~\ref{fig:app1}(b), where
\begin{equation}
\begin{gathered}
\mathcal{C} =\cosh (\alpha+\beta-\xi) \cos \eta+1, \quad 
\mathcal{S}=\sinh (\alpha+\beta-\xi) \sin \eta, \\
\mathcal{C}_{\mathrm{i}}=\cosh \beta \cos \eta+1, \quad  \mathcal{S}_{\mathrm{i}}=\sinh \beta \sin \eta, ~~\\
\Theta =\tan ^{-1}\left(\frac{\mathcal{S}}{\mathcal{C}}\right) \in[0,2 \pi],\quad  \Theta_{\mathrm{i}}=\tan ^{-1}\left(\frac{\mathcal{S}_{\mathrm{i}}}{\mathcal{C}_{\mathrm{i}}}\right) .
\end{gathered}
\end{equation}
In addition, for notational convenience, we adopt the finite-difference operators $\delta_i$ and $\delta_j$ following 
\citet[Chapter 3]{KajishimaTaira2017CFD}
\begin{equation}
\begin{aligned}
& \left.\delta_{i}(q)\right|_{i, j}=q_{i+1 / 2, j}-q_{i-1 / 2, j},\left.\quad \delta_{j}(q)\right|_{i, j}=q_{i, j+1 / 2}-q_{i, j-1 / 2}, \\
& \left.\delta_{i} \delta_{j}(q)\right|_{i, j}=q_{i+1 / 2, j+1 / 2}-q_{i+1 / 2, j-1 / 2}-q_{i-1 / 2, j+1 / 2}+q_{i-1 / 2, j-1 / 2} ,
\end{aligned}
\label{eq:diff_ope}
\end{equation}
where the indices $i$ and $j$ denote discrete grid points in the $\xi$- and $\eta$-directions, respectively.
Using these definitions, the physical grid spacings
\begin{equation}
\begin{alignedat}{2}
  \left(\Gamma_{\xi}\right)_{v}&=\left(h_{\xi}\delta_i(\xi)\right)_{v}=\delta_{i}\left(g_{\xi}\right),&\quad
  \left(\Gamma_{\eta}\right)_{u}&=\left(h_{\eta}\delta_j(\eta)\right)_{u}=\delta_{j}\left(g_{\eta}\right),\\
  \left(\Gamma_{\xi}\right)_{u}&=\left(h_{\xi}\,\delta_i(\xi)\right)_{u}
  =\frac{\left(\Gamma_{\eta}\right)_{u}{\overline{\delta_{i}(\xi)}}^{\,i}}{\delta_{j}(\eta)},&\quad
  \left(\Gamma_{\eta}\right)_{v}&=\left(h_{\eta}\,\delta_j(\eta)\right)_{v}
  =\frac{\left(\Gamma_{\xi}\right)_{v}{\overline{\delta_{j}(\eta)}}^{\,j}}{\delta_{i}(\xi)},
  \label{eq:Gamma_defs}
\end{alignedat}
\end{equation}
and the cell areas
\begin{equation}
\begin{gathered}
\left(\Gamma\right)_{p}=\left(h\,\delta_i(\xi)\,\delta_j(\eta)\right)_{p}=\delta_{i}\delta_{j}(g),~~
\left(\Gamma\right)_{u}=\left(\Gamma_{\xi}\right)_u \left(\Gamma_{\eta}\right)_u,~~
\left(\Gamma\right)_{v}=\left(\Gamma_{\xi}\right)_v
\left(\Gamma_{\eta}\right)_v,
\end{gathered}
\end{equation}
are evaluated.
The subscripts $u$ and $v$ indicate the locations at which $u_{\xi}$ and $u_{\eta}$ are defined, respectively, while $p$ denotes the cell centre. 
A superscript overbar represents the interpolation operation
\begin{equation}
  \left.\overline{q}^{\,i}\right|_{i,j}=\frac{q_{i+1/2,j}+q_{i-1/2,j}}{2},\qquad
  \left.\overline{q}^{\,j}\right|_{i,j}=\frac{q_{i,j+1/2}+q_{i,j-1/2}}{2}.
\end{equation}

The continuity equation~\eqref{eq:cnt} is discretised in finite-difference form as
\begin{equation}
\left.\nabla\cdot\bm{u}\right|_{i,j}=\frac{\left.\delta_i(U_\xi)\right|_{i,j}+\left.\delta_j(U_\eta)\right|_{i,j}}{\left.(\Gamma)_p\right|_{i,j}},\quad
  U_\xi =(\Gamma_\eta)_u u_\xi,\quad
  U_\eta=(\Gamma_\xi)_v u_\eta,
  \label{eq:d_cnt}
\end{equation}
where $U_{\xi}$ and $U_{\eta}$ denote the contravariant velocity components.
For the spatial discretisation of the momentum equation~\eqref{eq:ns}, the terms evaluated on the cell faces are written as
\begin{equation}
\begin{aligned}
\left.\frac{\partial u_\xi}{\partial t}\right|_{i+1/2,j}
&=-\left.\mathcal{A}_\xi\right|_{i+1/2,j}-\left.\mathcal{P}_\xi\right|_{i+1/2,j}+\left.\operatorname{Re}^{-1}\mathcal{V}_\xi\right|_{i+1/2,j},\\
\left.\frac{\partial u_\eta}{\partial t}\right|_{i,j+1/2}
&=-\left.\mathcal{A}_\eta\right|_{i,j+1/2}-\left.\mathcal{P}_\eta \right|_{i,j+1/2}+\left.\operatorname{Re}^{-1}\mathcal{V}_\eta\right|_{i,j+1/2},
\end{aligned}
\end{equation}
where $\mathcal{A}$, $\mathcal{P}$, and $\mathcal{V}$ denote the advection, pressure-gradient, and viscous terms, respectively.
The advection term is evaluated in the kinetic-energy-conserving form
${(\bm{u}^{\prime}\cdot\nabla)\bm{u}+\nabla\cdot(\bm{u}^{\prime}\bm{u})}/2$.
The pressure-gradient term is discretised as
\begin{equation}
  \left.\mathcal{P}_\xi\right|_{i+1 / 2, j}=  \frac{\left.\delta_{i}(p)\right|_{i+1 / 2, j}}{\left.\left(\Gamma_{\xi}\right)_{u}\right|_{i+1 / 2, j}}, \quad  
  \left.\mathcal{P}_\eta\right|_{i, j+1 / 2}=  \frac{\left.\delta_{j}(p)\right|_{i, j+1 / 2}}{\left.\left(\Gamma_{\eta}\right)_{v}\right|_{i, j+1 / 2}} .
  \label{eq:d_prs}
\end{equation}
The components $\mathcal{A}_{\xi}$, $\mathcal{A}_{\eta}$ and $\mathcal{V}_{\xi}$, $\mathcal{V}_{\eta}$ are discretised as
\begin{align}
\left.\mathcal{A}_{\xi}\right|_{i+1/2,j}&= \frac{1}{{2 (\Gamma)_u}|_{i+1/2, j}}
  \left\{ 
      \widetilde{\overline{{U}^{\prime}_{\xi}}^i \delta_i (u_{\xi})}^i
    + \overline{\widetilde{{U}^{\prime}_{\eta}}^i \delta_j (u_{\xi})}^j
    + \tilde{\delta}_i (\widetilde{{U}^{\prime}_{\xi}}^i \widetilde{u_{\xi}}^i) 
  \right. \notag \\ 
  & \quad \quad \left. 
    + \delta_j (\widetilde{{U}^{\prime}_{\eta}}^i \overline{u_{\xi}}^j)
    + 2\widetilde{ {u}^{\prime}_{\xi} \overline{u_{\eta}}^j \delta_j ((\Gamma_{\xi})_v)}^i
    - 2\widetilde{ \overline{{u}^{\prime}_{\eta} u_{\eta}}^j \delta_i ((\Gamma_{\eta})_u)}^i
  \right\}_{i+1/2, j}, \\
\left.\mathcal{A}_{\eta}\right|_{i,j+1/2}&= \frac{1}{{2 (\Gamma)_v}|_{i, j+1/2}}
  \left\{ 
      \overline{\overline{{U}^{\prime}_{\xi}}^j \delta_i (u_{\eta})}^i
    + \overline{\overline{{U}^{\prime}_{\eta}}^j \delta_j (u_{\eta})}^j
    + \delta_i (\overline{{U}^{\prime}_{\xi}}^j \overline{u_{\eta}}^i) 
  \right. \notag \\ 
  & \quad \quad\left. 
    + \delta_j (\overline{{U}^{\prime}_{\eta}}^j \overline{u_{\eta}}^j) 
    + 2{u}^{\prime}_{\eta} \overline{  \overline{u_{\xi}}^i \delta_i ((\Gamma_{\eta})_u)}^j
    - 2\overline{ \overline{{u}^{\prime}_{\xi} u_{\xi}}^j \delta_j ((\Gamma_{\xi})_v)}^j
  \right\}_{i, j+1/2},
\end{align}
\begin{align}
\left.\mathcal{V}_{\xi}\right|_{i+1/2,j}&= \frac{1}{\left.(\Gamma)_u\right|_{i+1 / 2, j}} 
  \left\{ 
    \delta_i\left({\overline{\left(\Gamma_\eta\right)_u}}^i \tau_{\xi \xi}\right)
    + \delta_j\left({\overline{\left(\Gamma_\xi\right)_u}}^j \tau_{\xi \eta}\right) 
    \right. \notag \\
    & \quad \quad \quad \quad \quad \quad \quad \quad \quad \quad \left.
    + \overline{\left(\delta_j\left(\left(\Gamma_{\xi}\right)_u\right) \tau_{\xi \eta}\right)}^j
    - \widetilde{ \delta_i ((\Gamma_{\eta})_u)\tau_{\eta \eta} }^i
  \right\}_{i+1/2,j}, \\
\left.\mathcal{V}_{\eta}\right|_{i,j+1/2}&=\frac{1}{\left.(\Gamma)_{v}\right|_{i, j+1 / 2}}
  \left\{
    \delta_{i}\left({\overline{\left(\Gamma_{\eta}\right)_{v}}}^{i} \tau_{\xi \eta}\right)
    +\delta_{j}\left({\overline{\left(\Gamma_{\xi}\right)_{v}}}^{j} \tau_{\eta \eta}\right)
  \right. \notag \\ 
    & \quad \quad \quad \quad \quad \quad \quad \quad \quad \quad \left. 
    +{\overline{\left(\delta_{i}\left(\left(\Gamma_{\eta}\right)_{v}\right) \tau_{\xi \eta}\right)}}^{i}
    -{\overline{\delta_{j}\left(\left(\Gamma_{\xi}\right)_{v}\right) \tau_{\xi \xi}}}^{j}
  \right\}_{i, j+1 / 2} ,
\end{align}
where the tilde operators $\widetilde{(\cdot)^{i}}$ and $\widetilde{\delta}^{i}$ represent the interpolation and difference operations applied at the boundary, defined as
\begin{equation}
\begin{alignedat}{2}
&\widetilde{f}^{\,i}|_{i+1 / 2, j}=
  \begin{cases}
    f|_{1, j} & i=0, \\
    f|_{N_{i}, j} & i=N_{i}, \\
    \bar{f}^{i}|_{i+1 / 2, j} & \text { otherwise. }
  \end{cases}&~~
&\widetilde{f}^{\,i}|_{i+1 / 2, j+1 / 2}=
  \begin{cases}
    f|_{1, j+1 / 2} & i=0, \\
    f|_{N_{i}, j+1 / 2} & i=N_{i}, \\
    \bar{f}^{i}|_{i+1 / 2, j+1 / 2} & \text { otherwise. }
  \end{cases}   \\
&\widetilde{f}^{\,i}|_{i,j}=
  \begin{cases}
    f|_{1/2,j} & i=1/2,\\
    f|_{N_i+1/2,j} & i=N_i+1/2,\\
    \overline{f}^{\,i}|_{i,j} & \text{otherwise}.
  \end{cases} &~~
&\widetilde{\delta}^{\,i}(f)|_{i+1/2,j}=
  \begin{cases}
    2(f|_{1,j}-f|_{1/2,j}) & i=0,\\
    2(f|_{N_i+1/2,j}-f|_{N_i,j}) & i=N_i,\\
    \delta_i(f)|_{i+1/2,j} & \text{otherwise}.
  \end{cases}
\end{alignedat}
\end{equation}
We next describe the boundary conditions.
The periodic boundary conditions in the $\eta$-direction, corresponding to \eqref{eq:bci} and \eqref{eq:bco}, are discretised as
\begin{equation}
\begin{alignedat}{1}
&\left.\phi\right|_{i,j\pm N_j} = \left.\phi\right|_{i,j},\qquad
\left.\phi\right|_{i,j\pm N_j+1/2} = \left.\phi\right|_{i,j+1/2},\\
&\left.\phi\right|_{i+1/2,j\pm N_j} = \left.\phi\right|_{i+1/2,j},\qquad
\left.\phi\right|_{i+1/2,j\pm N_j+1/2} = \left.\phi\right|_{i+1/2,j+1/2}.
\end{alignedat}
\label{eq:bc_periodic}
\end{equation}
On the inner cylinder surface, the kinematic and no-slip conditions are discretised as
\begin{equation}
\begin{alignedat}{1}
&\left.u_{\xi}\right|_{1/2,j}=\varepsilon \Omega \,\left.a_{y\xi}\right|_{1/2,j},\qquad
\left.u_{\xi}\right|_{N_i+1/2,j}=0,\\
&\left.\overline{u_{\eta}}^i\right|_{1/2,j+1/2}=R_{\mathrm{i}}\omega + \varepsilon \Omega \,\left.\overline{a_{y\eta}}^i\right|_{1/2,j+1/2},\qquad
\left.\overline{u_{\eta}}^i\right|_{N_i+1/2,j+1/2}=0,
\end{alignedat}
\end{equation}
where
\begin{equation}
  (a_{y\xi})_u = \left(\bm{e}_y \cdot \bm{e}_{\xi}\right)_{u} = -\frac{\delta_j\,((x)_a)}{(\Gamma_\eta)_{u}}, \qquad 
  (a_{y\eta})_v=\left(\bm{e}_y \cdot \bm{e}_{\eta}\right)_{v} = \frac{\delta_i\,((x)_a)}{(\Gamma_\xi)_{v}}.
\end{equation}

Finally, the components of the hydrodynamic force in \eqref{eq:frc_s} are expressed as
\begin{align}
F_x&=
  -\sum_{j=1}^{N_j}\left\{
    (\Gamma_\eta)_u (a_{x\xi})_u \overline{p}^i\right\}_{1/2,j} \notag\\
  & \quad +\operatorname{Re}^{-1}\sum_{j=1}^{N_j}\left\{(a_{x\xi})_u\overline{\overline{(\Gamma_{\eta})_u}^i\tau_{\xi\xi}}^i\right\}_{1/2,j}
  +\operatorname{Re}^{-1}\sum_{j=1}^{N_j}\left\{\overline{(a_{x\eta})_v}^i\overline{(\Gamma_\eta)_v}^i\tau_{\xi\eta}\right\}_{1/2,j+1/2}
    , \\
F_y&=
  -\sum_{j=1}^{N_j}\left\{
     (\Gamma_\eta)_u (a_{y\xi})_u \overline{p}^i\right\}_{1/2,j} \notag\\
  & \quad +\operatorname{Re}^{-1}\sum_{j=1}^{N_j}\left\{(a_{y\xi})_u\overline{\overline{(\Gamma_{\eta})_u}^i\tau_{\xi\xi}}^i\right\}_{1/2,j}
  +\operatorname{Re}^{-1}\sum_{j=1}^{N_j}\left\{\overline{(a_{y\eta})_v}^i\overline{(\Gamma_\eta)_v}^i\tau_{\xi\eta}\right\}_{1/2,j+1/2},
\end{align}
where
\begin{equation}
  (a_{x\xi})_u = \left(\bm{e}_x \cdot \bm{e}_{\xi}\right)_{u} = \frac{\delta_j\,((y)_a)}{(\Gamma_\eta)_{u}}, \qquad 
  (a_{x\eta})_v=\left(\bm{e}_x \cdot \bm{e}_{\eta}\right)_{v} = -\frac{\delta_i\,((y)_a)}{(\Gamma_\xi)_{v}}.
\end{equation}

\section{Direct solution procedure for the pressure Poisson equation}
\label{sec:appB}
In this section, we describe the solution procedure for the pressure Poisson equation in the SMAC (Simplified Marker and Cell) method.
In the SMAC method, the predicted velocity $\bm{u}^*$ is first computed, and then corrected by introducing the pressure correction $\phi$ so that the velocity field at the next time step satisfies the incompressibility condition.
In this case, $\phi$ satisfies the pressure Poisson equation shown in equation
\begin{equation}
  \nabla^2 \phi = \frac{\rho \nabla \cdot \bm{u}^*}{\Delta t},
  \label{eq:pois1}
\end{equation}
where $\Delta t$ denotes the time increment.
By applying a second-order finite-volume discretisation to equation
\begin{equation}
\frac{1}{(\Gamma)_{p}|_{i,j}}\left(
 \delta_i\left(\frac{\delta_j(\eta)\delta_i(\phi)}{\overline{\delta_i(\xi )}^i}\right)
+\delta_j\left(\frac{\delta_i(\xi )\delta_j(\phi)}{\overline{\delta_j(\eta)}^j}\right)
\right)_{i,j}
=D|_{i,j},\ \ i\in[1,N_i],\ j\in[1,N_j],
\label{eq:pois2}
\end{equation}
we obtain the discrete equations given by equations
\begin{equation}
D|_{i,j}=
\frac{\rho\{\delta_i(U_\xi)|_{i,j}+\delta_j(U_\eta)|_{i,j}\}}{(\Delta t)(\Gamma)_p|_{i,j}}.
\end{equation}
Since the present problem is periodic in the $\eta$ direction, periodic boundary conditions are imposed on $\phi$ and $D$, as given in equation
\begin{equation}
  \phi|_{i,j\pm N_j}=\phi|_{i,j},\ \ 
  D|_{i,j\pm N_j}=D|_{i,j}.
  \label{eq:bc_eta1}
\end{equation}
In addition, a Neumann boundary condition is imposed on $\phi$ at the boundaries in the $\xi$ direction, as given in equation
\begin{equation}
\phi|_{0,j}=\phi|_{1,j},\ \ \phi|_{N_i+1,j}=\phi|_{N_i,j}.
\label{eq:bc_phi1}
\end{equation}
Because the coordinates $\xi$ and $\eta$ depend only on $i$ and $j$, respectively, equation~\eqref{eq:pois2} can be rewritten as equation
\begin{equation}
\left\{
 \frac{1}{\delta_i(\xi )}\delta_i\left(\frac{\delta_i(\phi)}{\overline{\delta_i(\xi )}^i}\right)
+\frac{1}{\delta_j(\eta)}\delta_j\left(\frac{\delta_j(\phi)}{\overline{\delta_j(\eta)}^j}\right)\right\}_{i,j}
=S|_{i,j},\ \ i\in[1,N_i],\ j\in[1,N_j],
\label{eq:pois3}
\end{equation}
\begin{equation}
S|_{i,j}=\left\{\frac{(\Gamma_p) D}{\delta_i(\xi)\delta_j(\eta)}\right\}_{i,j},
\end{equation}
in which the terms associated with $\xi$ and $\eta$ are separated.
By expanding equation~\eqref{eq:pois3} with the periodic boundary condition~\eqref{eq:bc_eta1} and the Neumann condition~\eqref{eq:bc_phi1}, we obtain the five-point stencil difference equation for $\phi$, given by equation
\begin{equation}
\begin{split}&
 a_w(i)\phi|_{i-1,j}
+a_e(i)\phi|_{i+1,j}
-(a_w(i)+a_e(i))\phi|_{i,j}\\&
+b_s(j)\phi|_{i,(j+N_j-2)\%N_j+1}+b_n(j)\phi|_{i,(j+N_j)\%N_j+1}\\&
-(b_s(j)+b_n(j))\phi|_{i,j}
=S|_{i,j},\ \ i\in[1,N_i],\ j\in[1,N_j],
\label{eq:pois4}
\end{split}
\end{equation}
~\eqref{eq:pois4}.
where $a_w(i)$ and $a_e(i)$ are the coefficients associated with the finite difference in the $\xi$ direction, while $b_s(j)$ and $b_n(j)$ are those associated with the finite difference in the $\eta$ direction.
The symbol $\%$ denotes the modulo operation, which reflects the periodicity in the $\eta$ direction ($j=0\rightarrow N_j$, $j=N_j+1\rightarrow 1$).

In the following, based on the procedure of \citet{muller2019fft}, we describe a fast method for solving equation~\eqref{eq:pois4}.
Specifically, the second-order difference term in the $\eta$ direction in equation~\eqref{eq:pois4} is first written in matrix form as a linear operation acting on the discrete vector $\phi_i$ in the $\eta$ direction, and the corresponding coefficient matrix is defined as $\bm{A}$.
Next, by expanding $\phi_i$ in the eigenmodes of $\bm{A}$ and diagonalising the difference operator in the $\eta$ direction, the two-dimensional problem is reduced to a tridiagonal system for each eigenmode.
For this purpose, the $\eta$-direction difference term in equation~\eqref{eq:pois4} is written as $\sum_{J=1}^{N_j}A_{j,J}\phi|_{i,J}$, and the coefficient matrix $A_{j,J}$ 
\begin{equation}
A_{I,J}=\left\{
\begin{array}{ll}
b_s(I),&\ J=(I+N_j-2)\%N_j+1,\\
-b_s(I)-b_n(I),&\ J=(I+N_j-1)\%N_j+1,\\
b_n(I),&\ J=(I+N_j)\%N_j+1,\\
0,&\ \mbox{otherwise},
\end{array}
\right.
\end{equation}
is introduced as above.
Here, $A_{j,J}$ is the matrix representation of the second-order difference operator in the $\eta$ direction, and is defined so as to be non-zero only for $J=j-1$, $j$, and $j+1$.
Next, let the eigenvalues and eigenvectors of $\bm{A}$ be $\lambda^{(K)}$ and $\bm{v}^{(K)}$ ($K=1,\ldots,N_j$), respectively, and expand $\phi_i$ in these eigenmodes.
Then, the $\eta$-direction difference term is separated mode by mode, as shown in equation
\begin{equation}
\sum_{J=1}^{N_j}A_{j,J}\phi|_{i,J}=\sum_{K=1}^{N_j}\sum_{J=1}^{N_j}A_{j,J}v_{J}^{(K)}\hat{\phi}_{i}^{(K)}
=\sum_{K=1}^{N_j}\lambda^{(K)}v_{j}^{(K)}\hat{\phi}_{i}^{(K)},
\label{eq:pois6}
\end{equation}
\begin{equation}
\hat{\phi}_{i}^{(K)}=\sum_{j=1}^{N_j}(v^{-1})_j^{(K)}\phi|_{i,j}.
\end{equation}
Therefore, by applying the inverse transform to both sides of equation~\eqref{eq:pois4}, we obtain, for each mode $K$, a one-dimensional linear system in the $\xi$ direction, given by equation
\begin{equation}
 a_w(i)\hat{\phi}_{i-1}^{(K)}
+a_e(i)\hat{\phi}_{i+1}^{(K)}
+\left\{-(a_w(i)+a_e(i))+\lambda^{(K)}\right\}\hat{\phi}_{i}^{(K)}=\hat{S}_{i}^{(K)},\ \ i\in[1,N_i],\ K\in[1,N_j].
\label{eq:pois7}
\end{equation}
\begin{equation}
\hat{S}_{i}^{(K)}=\sum_{j=1}^{N_j}(v^{-1})_j^{(K)}S|_{i,j}.
\label{eq:hatS}
\end{equation}
~\eqref{eq:pois7}.
For each $K$, equation~\eqref{eq:pois7} is a tridiagonal linear system in the $i$ direction, and can therefore be solved efficiently by the TDMA method.
Finally, the solution $\phi|_{i,j}$ ($i\in[1,N_i]$, $j\in[1,N_j]$) is obtained by the inverse transformation shown in equation
\begin{equation}
\phi|_{i,j}=\sum_{K=1}^{N_j}v_{j}^{(K)}\hat{\phi}_{i}^{(K)}.
\label{eq:phi}
\end{equation}

\section{Drag acting on the inner cylinder in the limit of a power-law index close to unity}
\label{sec:app_asymp}
In this section, we derive an approximate expression for the drag force $F_y$ acting on the inner cylinder when the power-law index $n$ is very close to unity, that is, when the fluid exhibits only weak non-Newtonian behaviour.
We define a small parameter representing the deviation of the power-law index $n$ from unity as $\epsilon = 1-n~(0< \epsilon \ll 1)~$.
Based on the method of perturbation, we write the velocity $\bm{u}$, pressure $p$, viscosity $\mu$, and stress tensor $\bm{\sigma}$ in an expansion form:
\begin{subequations}
\begin{align}
\bm{u} &= \bm{u}^{(0)} + \epsilon \bm{u}^{(1)} + O(\epsilon^2), \\
{p} &= {p}^{(0)} + \epsilon {p}^{(1)} + O(\epsilon^2), \\
\operatorname{Re}^{-1}\mu &= \mu^{(0)} + \epsilon \mu^{(1)} + O(\epsilon^2), \\
\bm{\sigma} &= \bm{\sigma}^{(0)} + \epsilon \bm{\sigma}^{(1)} + O(\epsilon^2).
\label{eq:ep_exp}
\end{align}
\end{subequations}
Applying the Taylor expansion to the viscosity $\mu$ with respect to $\epsilon$ gives
\begin{equation}
 \mu^{(0)}=1, \quad \mu^{(1)} = -\ln (2\bm{S}^{(0)}\colon\bm{S}^{(0)})^{1/2}. 
\end{equation}
The governing equations at $O(\epsilon^0)$ are given by
\begin{subequations}
\begin{align}
&\nabla\cdot\bm{u}^{(0)} = 0, \quad \nabla\cdot\bm{\sigma}^{(0)} = 0, \numtag{a,b}\label{eq:ep0}\\
&\bm{\sigma}^{(0)} = -p^{(0)}\bm{I} + 2\bm{S}^{(0)}, \numtag{c}\label{eq:ep0_sigma}\\
&\bm{u}^{(0)} =
\begin{cases}
\varepsilon \Omega  \bm{e}_y + \bm{\omega}\times (\bm{x}-\bm{x}_\mathrm{i}) & \text{on } S_{\mathrm{i}},\\
\bm{0} & \text{on } S_{\mathrm{o}}.\numtag{d}
\end{cases}
\label{eq:ep0_bc}
\end{align}
\end{subequations}
These correspond to the solution for a Newtonian fluid.
The governing equations at $O(\epsilon^1)$ are
\begin{subequations}
\begin{align}
&\nabla \cdot \bm{u}^{(1)} = 0, \quad \nabla \cdot \bm{\sigma}^{(1)} = 0, \numtag{a,b}\label{eq:ep1}\\
&\bm{\sigma}^{(1)} = -p^{(1)}\bm{I} + 2\left(\bm{S}^{(1)} + \mu^{(1)}\bm{S}^{(0)}\right), \numtag{c}\label{eq:ep1_sigma} \\ 
&\bm{u}^{(1)} = \bm{0} ~~~ \text{on } S_{\mathrm{i}} \cup S_{\mathrm{o}}, \numtag{d}\label{eq:ep1_bc}
\end{align}
\end{subequations}
The drag force $F_y$ acting on the inner cylinder is expanded from \eqref{eq:ep_exp} as
\begin{equation}
\operatorname{Re}F_y = F_y^{(0)} + \epsilon F_y^{(1)} + O(\epsilon^2),
\end{equation}
where $F_y^{(0)} = -\int_{S_{\mathrm{i}}} \bm{n} \cdot (\bm{e}_y \cdot \bm{\sigma}^{(0)}) ~ \mathrm{d} S$ and $F_y^{(1)} = -\int_{S_{\mathrm{i}}} \bm{n} \cdot (\bm{e}_y \cdot \bm{\sigma}^{(1)}) ~ \mathrm{d} S$.
In general, the evaluation of $F_y^{(1)}$ requires solving the flow field at $O(\epsilon^1)$.
However, in this section, by using the Lorentz reciprocal theorem, we derive $F_y^{(1)}$ only from the known $O(\epsilon^0)$ field and an auxiliary field, without directly solving the first-order flow field.
For this purpose, we introduce an auxiliary Newtonian field $(\hat{\bm{u}},\hat{\bm{\sigma}})$:
\begin{subequations}
\begin{align}
&\nabla \cdot \hat{\bm{u}} = 0,\quad \nabla \cdot \hat{\bm{\sigma}} = 0, \numtag{a,b}\label{eq:aux}\\
&\hat{\bm{\sigma}} = -\hat{p}\bm{I} + 2\hat{\bm{S}},
\numtag{c}\label{eq:aux_sigma}\\
&\hat{\bm{u}} = 
\begin{cases}
\bm{e}_y &\text{on } S_{\mathrm{i}}, \\
\bm{0}   &\text{on } S_{\mathrm{o}}. \numtag{d}
\label{eq:aux_bc}
\end{cases}
\end{align}
\end{subequations}
This is the flow field around an inner cylinder translating in the $y$ direction.
Taking the inner products of the momentum equations \eqref{eq:ep1} and \eqref{eq:aux} with $\hat{\bm{u}}$ and $\bm{u}^{(1)}$, respectively, subtracting them, and integrating over the fluid domain $V$, we obtain
\begin{equation}
\int_V \nabla \cdot \left(
  \bm{u}^{(1)} \cdot \hat{\bm{\sigma}}
- \hat{\bm{u}} \cdot \bm{\sigma}^{(1)}\right) ~ \mathrm{d} V
=
\int_V \left(
  \hat{\bm{\sigma}} \colon \nabla \bm{u}^{(1)}
- \bm{\sigma}^{(1)} \colon \nabla \hat{\bm{u}}  \right)  ~ \mathrm{d}V.
\end{equation}
The volume integral on the left-hand side can be converted into surface integrals over the boundaries $S_\mathrm{i}$ and $S_\mathrm{o}$ by the Gauss divergence theorem.
Using the boundary conditions \eqref{eq:ep1_bc} and \eqref{eq:aux_bc}, the first term on the left-hand side vanishes, and the second term coincides with the correction term $F_y^{(1)}$ of the drag force.
On the other hand, the integrand on the right-hand side can be rearranged by substituting the constitutive equation \eqref{eq:ep1_sigma} and using the incompressibility condition and the symmetry of the strain-rate tensor ($ \hat{\bm{S}} \colon \nabla \bm{u}^{(1)} =  {\bm{S}}^{(1)} \colon \nabla \hat{\bm{u}}$), yielding
\begin{equation}
  \hat{\bm{\sigma}} \colon \nabla \bm{u}^{(1)} -\bm{\sigma}^{(1)} \colon \nabla \hat{\bm{u}}= -2 \mu^{(1)} \bm{S}^{(0)} \colon \hat{\bm{S}}.
\end{equation}
Therefore, $F_y^{(1)}$ can be written in the volume-integral form
\begin{equation}
 F_y^{(1)} = -\int_V 2 \mu^{(1)} \bm{S}^{(0)} \colon \hat{\bm{S}}~ \mathrm{d} V.
\end{equation}
Hence, the $O(\epsilon)$ approximation for the drag force $F_y$ acting on the inner cylinder is expressed, using only the known Newtonian flow field $\bm{S}^{(0)}$ and the auxiliary field $\hat{\bm{S}}$, as
\begin{equation}F_y \approx \frac{1}{\operatorname{Re}}\int_V 2 \left[ - 1 + \epsilon \ln(2 \bm{S}^{(0)} \colon \bm{S}^{(0)})^{\frac{1}{2}} \right] \bm{S}^{(0)} \colon \hat{\bm{S}} ~ \mathrm{d} V.\end{equation}

\bibliographystyle{jfm}
\bibliography{jfm}

@article{cross1965rheology,
  title={{Rheology of non-Newtonian fluids: a new flow equation for pseudoplastic systems}},
  author={Cross, M. M.},
  journal={J. Colloid Interface Sci.},
  volume={20},
  number={5},
  pages={417--437},
  year={1965},
  publisher={Elsevier}
}

@article{carreau1972rheological,
  title={Rheological equations from molecular network theories},
  author={Carreau, P.J.},
  journal={Trans. Soc. Rheol.},
  volume={16},
  number={1},
  pages={99--127},
  year={1972},
  publisher={The Society of Rheology}
}

@book{chhabra2023bubbles,
  title={{Bubbles, Drops, and Particles in Non-Newtonian Fluids}},
  author={Chhabra, R. P. and Patel, S. A.},
  year={2023},
  publisher={CRC press}
}

@article{amsden1970simplified,
  title={{A simplified MAC technique for incompressible fluid flow calculations}},
  author={Amsden, A. A. and Harlow, F. H.},
  journal={J. Comput. Phys.},
  volume={6},
  number={2},
  pages={322--325},
  year={1970},
  publisher={Elsevier}
}

@article{zhang2023drag,
  title={Drag force on an oscillatory spherical bubble in shear-thinning fluid},
  author={Zhang, X. and Sugiyama, K. and Watamura, T.},
  journal={J. Fluid Mech.},
  volume={959},
  pages={A3},
  year={2023},
  publisher={Cambridge University Press}
}

@article{brindley1979flow,
    author = {Brindley, J. and Elliott, L. and McKay, J. T.},
    title = {Flow in a Whirling Rotor Bearing},
    journal = {J. Appl. Mech.},
    volume = {46},
    number = {4},
    pages = {767--771},
    year = {1979},
    month = {12},
    doi = {10.1115/1.3424651},
}

@article{podryabinkin2011moment,
  title={Moment and forces exerted on the inner cylinder in eccentric annular flow},
  author={Podryabinkin, E.V. and Rudyak, V. Y.},
  journal={J. Eng. Thermophys.},
  volume={20},
  number={3},
  pages={320--328},
  year={2011},
  publisher={Springer}
}

@article{kazakia1978flow,
  title={{Flow of a Newtonian fluid between eccentric rotating cylinders and related problems}},
  author={Kazakia, J.Y. and Rivlin, R.S.},
  journal={Stud. Appl. Math.},
  volume={58},
  number={3},
  pages={209--247},
  year={1978},
  publisher={Wiley Online Library}
}

@book{saffman1992vortex,
  title={{Vortex Dynamics}},
  author={Saffman, P.G.},
  year={1992},
  publisher={Cambridge University Press}
}

@book{wu2006vorticity,
  title={{Vorticity and Vortex Dynamics}},
  author={Wu, J. and Ma, H. and Zhou, M.},
  year={2006},
  publisher={Springer Science \& Business Media}
}

@incollection{von1935general,
  title={{General Aerodynamic Theory: Perfect Fluids}},
  booktitle={Aerodynamic Theory II},
  author={von K{\'a}rm{\'a}n, T. and Burgers, J.M.},
  year={1935},
  publisher={Springer}
}

@ARTICLE{Zhang2016Fundamentals_kw,
  title     = "{Fundamentals and applications of inertial microfluidics: a
               review}",
  author    = "Zhang, J. and Yan, S. and Yuan, D. and Alici, G. and
               Nguyen, N. T. and Warkiani, M.E. and Li, W.",
  journal   = "Lab Chip",
  publisher = "pubs.rsc.org",
  volume    =  16,
  number    =  1,
  pages     = "10--34",
  year      =  2016,
  language  = "en"
}

@ARTICLE{DiCarlo2009Inertial_ed,
  title     = "{Inertial microfluidics}",
  author    = "Di Carlo, D.",
  journal   = "Lab Chip",
  publisher = "pubs.rsc.org",
  volume    =  9,
  number    =  21,
  pages     = "3038--3046",
  year      =  2009,
  language  = "en"
}

@ARTICLE{Zhou2013Fundamentals_yu,
  title    = "{Fundamentals of inertial focusing in microchannels}",
  author   = "Zhou, J. and Papautsky, I.",
  journal  = "Lab Chip",
  volume   =  13,
  number   =  6,
  pages    = "1121--1132",
  year     =  2013,
  language = "en"
}

@article{newkirk1925shaft,
  title={Shaft whipping due to oil action in journal bearings},
  author={Newkirk, B.L. and Taylor, H.D.},
  journal={General Electric Review},
  volume={28},
  number={8},
  pages={559--568},
  year={1925}
}

@article{benckert1980flow,
  title={Flow induced spring coefficients of labyrinth seals for application in rotor dynamics},
  author={Benckert, H. and Wachter, J.},
  journal={NASA. Lewis Res. Center Rotodyn. Instability Probl. in High-Performance Turbomachinery},
  year={1980}
}

@article{muszynska1986whirl,
  title={Whirl and whip-rotor/bearing stability problems},
  author={Muszynska, A.},
  journal={J. Sound Vib.},
  volume={110},
  number={3},
  pages={443--462},
  year={1986},
  publisher={Elsevier}
}

@article{sinha2017review,
  title={A review on particulate slurry erosive wear of industrial materials: In context with pipeline transportation of mineral--slurry},
  author={Sinha, S.L. and Dewangan, S.K. and Sharma, A.},
  journal={Part. Sci. Technol.},
  volume={35},
  number={1},
  pages={103--118},
  year={2017},
  publisher={Taylor \& Francis}
}

@article{alam2016slurry,
  title={Slurry erosion of pipeline steel: effect of velocity and microstructure},
  author={Alam, T. and Aminul I. M. and Farhat, Z.N.},
  journal={J. Tribol.},
  volume={138},
  number={2},
  pages={021604},
  year={2016},
  publisher={American Society of Mechanical Engineers}
}

@article{zhang2016study,
  title={Study on erosion wear of fracturing pipeline under the action of multiphase flow in oil \& gas industry},
  author={Zhang, J. and Kang, J. and Fan, J. and Gao, J.},
  journal={J. Nat. Gas Sci. Eng.},
  volume={32},
  pages={334--346},
  year={2016},
  publisher={Elsevier}
}

@article{bretherton1962motion,
  title={{The motion of rigid particles in a shear flow at low Reynolds number}},
  author={Bretherton, F.P.},
  journal={J. Fluid Mech.},
  volume={14},
  number={2},
  pages={284--304},
  year={1962},
  publisher={Cambridge University Press}
}

@article{ho1974inertial,
  title={Inertial migration of rigid spheres in two-dimensional unidirectional flows},
  author={Ho, B.P. and Leal, L.G.},
  journal={J. Fluid Mech.},
  volume={65},
  number={2},
  pages={365--400},
  year={1974},
  publisher={Cambridge University Press}
}

@article{vasseur1976lateral,
  title={The lateral migration of a spherical particle in two-dimensional shear flows},
  author={Vasseur, P. and Cox, R.G.},
  journal={J. Fluid Mech.},
  volume={78},
  number={2},
  pages={385--413},
  year={1976},
  publisher={Cambridge University Press}
}

@article{kamal1966separation,
    author = {Kamal, M. M.},
    title = {Separation in the Flow Between Eccentric Rotating Cylinders},
    journal = {J. Basic Eng.},
    volume = {88},
    number = {4},
    pages = {717--724},
    year = {1966},
    doi = {10.1115/1.3645951},
}

@ARTICLE{Ballal1976Flow_ag,
  title     = "{Flow of a Newtonian fluid between eccentric rotating cylinders:
               Inertial effects}",
  author    = "Ballal, B. Y. and Rivlin, R. S.",
  journal   = "Arch. Ration. Mech. Anal.",
  publisher = "Springer Nature",
  volume    =  62,
  number    =  3,
  pages     = "237--294",
  year      =  1976,
  language  = "en"
}

@ARTICLE{DiPrima1972Flow_cp,
  title     = "{Flow between eccentric rotating cylinders}",
  author    = "DiPrima, R. C. and Stuart, J. T.",
  journal   = "J. Lubr. Technol.",
  publisher = "ASME International",
  volume    =  94,
  number    =  3,
  pages     = "266--274",
  year      =  1972,
  language  = "en"
}

@ARTICLE{Brennen1976Flow_wd,
  title     = "{On the flow in an annulus surrounding a whirling cylinder}",
  author    = "Brennen, C.",
  journal   = "J. Fluid Mech.",
  publisher = "Cambridge University Press (CUP)",
  volume    =  75,
  number    =  1,
  pages     = "173--191",
  year      =  1976,
  language  = "en"
}

@article{shi2020lift,
  title={Lift forces on solid spherical particles in wall-bounded flows},
  author={Shi, P. and Rzehak, R.},
  journal={Chem. Eng. Sci.},
  volume={211},
  pages={115264},
  year={2020},
  publisher={Elsevier}
}

@ARTICLE{SanAndres1984Flow_nl,
  title     = "{Flow between eccentric rotating cylinders}",
  author    = "San, A.A. and Szeri, A.Z.",
  journal   = "J. Appl. Mech.",
  publisher = "ASME International",
  volume    =  51,
  number    =  4,
  pages     = "869--878",
  year      =  1984,
  language  = "en"
}

@article{yamada1968flow,
  title={On the flow between eccentric rotating cylinders when the outer cylinder rotates},
  author={Yamada, Y. and Nakabayashi, K.},
  journal={Bull. JSME},
  volume={11},
  number={45},
  pages={455--462},
  year={1968},
  publisher={The Japan Society of Mechanical Engineers}
}

@ARTICLE{Feng2007Orbital_ns,
  title     = "{On the orbital motion of a rotating inner cylinder in annular
               flow}",
  author    = "Feng, S. and Li, Q. and Fu, S.",
  journal   = "Int. J. Numer. Methods Fluids",
  publisher = "Wiley",
  volume    =  54,
  number    =  2,
  pages     = "155--173",
  year      =  2007,
  language  = "en"
}

@article{resell2025fluid,
  title={Fluid forces in eccentric annular geometries with rotating and orbiting inner cylinder},
  author={Resell, {\AA}.A. and Giljarhus, K.E.T. and Mihai, R. and Skadsem, H.J.},
  journal={Phys. Fluids},
  volume={37},
  number={4},
  pages={047158},
  year={2025},
  publisher={AIP Publishing}
}

@ARTICLE{Magnaudet2011Reciprocal_ue,
  title     = "{A `reciprocal' theorem for the prediction of loads on a body
               moving in an inhomogeneous flow at arbitrary Reynolds number}",
  author    = "Magnaudet, J",
  journal   = "J. Fluid Mech.",
  publisher = "cambridge.org",
  volume    =  689,
  pages     = "564--604",
  year      =  2011
}

@ARTICLE{Masoud2019Reciprocal_pf,
  title     = "{The reciprocal theorem in fluid dynamics and transport
               phenomena}",
  author    = "Masoud, H. and Stone, H.A.",
  journal   = "J. Fluid Mech.",
  publisher = "Cambridge University Press (CUP)",
  volume    =  879,
  pages    = "P1",
  year      =  2019,
  language  = "en"
}

@ARTICLE{Wu2007Integral_bg,
  title     = "{Integral force acting on a body due to local flow structures}",
  author    = "Wu, J.-Z. and Lu, X.-Y. and Zhuang, L.-X.",
  journal   = "J. Fluid Mech.",
  publisher = "Cambridge University Press (CUP)",
  volume    =  576,
  pages     = "265--286",
  year      =  2007,
  language  = "en"
}

@ARTICLE{Wang2022Vortex_hw,
  title     = "{Vortex force map for incompressible multi-body flows with
               application to wing–flap configurations}",
  author    = "Wang, Y. and Zhao, X. and Graham, M. and Li, J.",
  journal   = "J. Fluid Mech.",
  publisher = "Cambridge University Press (CUP)",
  volume    =  953,
  pages     = "A37",
  year      =  2022,
  language  = "en"
}

@article{prakhar2025vortices,
  title={From vortices to forces--a data-driven framework for unsteady lift generation in three-dimensional vortex-dominated flows},
  author={Prakhar, S. and Seo, J.H. and Mittal, R.},
  journal={J. Fluid Mech.},
  volume={1020},
  pages={A23},
  year={2025},
  publisher={Cambridge University Press}
}

@ARTICLE{Noca1997Measuring_xl,
  title     = "{Measuring instantaneous fluid dynamic forces on bodies, using
               only velocity fields and their derivatives}",
  author    = "Noca, F. and Shiels, D. and Jeon, D.",
  journal   = "J. Fluids Struct.",
  publisher = "Elsevier BV",
  volume    =  11,
  number    =  3,
  pages     = "345--350",
  year      =  1997,
  language  = "en"
}

@ARTICLE{Li2018Vortex_dp,
  title     = "{Vortex force map method for viscous flows of general airfoils}",
  author    = "Li, J. and Wu, Z.-N.",
  journal   = "J. Fluid Mech.",
  publisher = "Cambridge University Press (CUP)",
  volume    =  836,
  pages     = "145--166",
  year      =  2018,
  language  = "en"
}

@ARTICLE{Howe1995Force_gn,
  title   = "{On the force and moment on a body in an incompressible fluid, with
             application to rigid bodies and bubbles at high and low Reynolds
             numbers}",
  author  = "Howe, M. S.",
  journal = "Q. J. Mech. Appl. Math.",
  volume  =  48,
  number  =  3,
  pages   = "401--426",
  year    =  1995
}

@article{gao2025weighted,
  title={Weighted integral methods for fluid force diagnostics in incompressible flows},
  author={Gao, A.-K. and Xie, C. and Lu, X.-Y.},
  journal={J. Fluid Mech.},
  volume={1024},
  pages={A57},
  year={2025},
  publisher={Cambridge University Press}
}

@article{menon2021initiation,
  title={On the initiation and sustenance of flow-induced vibration of cylinders: insights from force partitioning},
  author={Menon, K. and Mittal, R.},
  journal={J. Fluid Mech.},
  volume={907},
  pages={A37},
  year={2021},
  publisher={Cambridge University Press}
}

@book{happel1973low,
  title={{Low Reynolds Number Hydrodynamics}},
  author={Happel, J. and Brenner, H.},
  year={1973},
  edition = {2nd},
  publisher={Martinus Nijhoff.}
}

@book{KajishimaTaira2017CFD,
  title     = {{Computational Fluid Dynamics: Incompressible Turbulent Flows}},
  author    = {Kajishima, T. and Taira, K.},
  publisher = {Springer International Publishing},
  address   = {Cham, Switzerland},
  year      = {2017},
  isbn      = {978-3-319-45304-0},
  doi       = {10.1007/978-3-319-45304-0},
  url       = {https://doi.org/10.1007/978-3-319-45304-0}
}

@book{batchelor1967introduction,
  title={{An Introduction to Fluid Dynamics}},
  author={Batchelor, G. K.},
  year={1967},
  publisher={Cambridge University Press}
}

@article{muller2019fft,
  title={{An FFT-based solution method for the Poisson equation on 3D spherical polar grids}},
  author={M{\"u}ller, B. and Chan, C.},
  journal={Astrophys. J.},
  volume={870},
  number={1},
  pages={43},
  year={2019},
  publisher={The American Astronomical Society}
}

@article{watamura2025transition,
  title={Transition of flow between eccentrically rotating double cylinders},
  author={Watamura, T. and Sugiyama, K. and Takagi, S.},
  journal={Int. J. Multiph. Flow.},
  volume={189},
  pages={105236},
  year={2025},
  publisher={Elsevier}
}

@article{wang2003lift,
  title={Lift forces on a cylindrical particle in plane Poiseuille flow of shear thinning fluids},
  author={Wang, J. and Joseph, D.D.},
  journal={Phys. Fluids},
  volume={15},
  number={8},
  pages={2267--2278},
  year={2003},
  publisher={American Institute of Physics}
}

@article{hazra2021dynamics,
  title={{Dynamics of rigid particles in a confined flow of viscoelastic and strongly shear-thinning fluid at very small Reynolds numbers}},
  author={Hazra, S. and Nath, A. and Mitra, S.K. and Sen, A.K.},
  journal={Phys. Fluids},
  volume={33},
  number={5},
  pages={052001},
  year={2021},
  publisher={AIP Publishing}
}

@article{li2015dynamics,
  title={Dynamics of particle migration in channel flow of viscoelastic fluids},
  author={Li, G. and McKinley, G.H. and Ardekani, A.M.},
  journal={J. Fluid Mech.},
  volume={785},
  pages={486--505},
  year={2015},
  publisher={Cambridge University Press}
}

@article{yamashita2025focusing,
  title={Focusing patterns of spherical particles in a viscoelastic flow through square channels},
  author={Yamashita, H. and Miki, Y. and Yoneyama, D. and Higashi, K. and Tsukasa, T. and Tagawa, Y. and Yokoyama, N. and Itano, T. and Sugihara-Seki, M.},
  journal={J. Fluid Mech.},
  volume={1022},
  pages={A44},
  year={2025},
  publisher={Cambridge University Press}
}

@article{raoufi2019experimental,
  title={Experimental and numerical study of elasto-inertial focusing in straight channels},
  author={Raoufi, M.A. and Mashhadian, A. and Niazmand, H. and Asadnia, M. and Razmjou, A. and Warkiani, M.E.},
  journal={Biomicrofluidics},
  volume={13},
  number={3},
  pages={034103},
  year={2019},
  publisher={AIP Publishing}
}

@article{hu2020influence,
  title={{Influence of non-Newtonian power law rheology on inertial migration of particles in channel flow}},
  author={Hu, X. and Lin, J. and Chen, D. and Ku, X.},
  journal={Biomicrofluidics},
  volume={14},
  number={1},
  pages = {014105},
  year={2020},
  publisher={AIP Publishing}
}

@article{szeri1995flow,
  title={Flow between finite, steadily rotating eccentric cylinders},
  author={Szeri, A.Z. and Al-Sharif, A.},
  journal={Theor. Comput. Fluid Dyn.},
  volume={7},
  number={1},
  pages={1--28},
  year={1995},
  publisher={Springer}
}

@article{volpi2024whirling,
  title={Whirling dynamics of a drill-string with fluid--structure interaction},
  author={Volpi, L.P. and Cayeux, E. and Time, R.W.},
  journal={Geoenergy Sci. Eng.},
  volume={232},
  pages={212423},
  year={2024},
  publisher={Elsevier}
}

@inproceedings{alderman1988rheological,
  title={The rheological properties of water-based drilling fluids},
  author={Alderman, N.J. and Ram Babu, D. and Hughes, T.L. and Maitland, G.C.},
  booktitle={Proc. Xth Int. Cong. Rheology, Sydney},
  pages={140--142},
  year={1988}
}

@article{berker1995effect,
  title={Effect of polymer on flow in journal bearings},
  author={Berker, A. and Bouldin, M.G. and Kleis, S.J. and VanArsdale, W.E.},
  journal={J. Non-Newton. Fluid Mech.},
  volume={56},
  number={3},
  pages={333--347},
  year={1995},
  publisher={Elsevier}
}

@article{carrasco2010non,
  title={{Non-Newtonian fluid displacements in horizontal narrow eccentric annuli: effects of slow motion of the inner cylinder}},
  author={Carrasco-Teja, M. and Frigaard, I.A.},
  journal={J. Fluid Mech.},
  volume={653},
  pages={137--173},
  year={2010},
  publisher={Cambridge University Press}
}

@article{birch2004force,
  title={{Force production and flow structure of the leading edge vortex on flapping wings at high and low Reynolds numbers}},
  author={Birch, J.M. and Dickson, W.B. and Dickinson, M.H.},
  journal={J. Exp. Biol.},
  volume={207},
  number={7},
  pages={1063--1072},
  year={2004},
  publisher={Company of Biologists}
}

@article{sane2003aerodynamics,
  title={The aerodynamics of insect flight},
  author={Sane, S.P.},
  journal={J. Exp. Biol.},
  volume={206},
  number={23},
  pages={4191--4208},
  year={2003},
  publisher={Company of Biologists}
}

@article{otomo2021unsteady,
  title={Unsteady lift on a high-amplitude pitching aerofoil},
  author={{\=O}tomo, S. and Henne, S. and Mulleners, K. and Ramesh, K. and Viola, I.M.},
  journal={Exp. Fluids.},
  volume={62},
  number={1},
  pages={6},
  year={2021},
  publisher={Springer}
}

@article{seo2022improved,
  title={Improved swimming performance in schooling fish via leading-edge vortex enhancement},
  author={Seo, J.H. and Mittal, R.},
  journal={Bioinspir. Biomim.},
  volume={17},
  number={6},
  pages={066020},
  year={2022},
  publisher={IOP Publishing}
}

@article{matsumoto1999vortex,
  title={Vortex shedding of bluff bodies: a review},
  author={Matsumoto, M.},
  journal={J. Fluids Struct.},
  volume={13},
  number={7-8},
  pages={791--811},
  year={1999},
  publisher={Elsevier}
}

@article{barnes1983vortex,
  title={Vortex shedding in unsteady flow},
  author={Barnes, F.H. and Grant, I.},
  journal={J. Wind Eng. Ind. Aerodyn.},
  volume={11},
  number={1-3},
  pages={335--344},
  year={1983},
  publisher={Elsevier}
}

\end{document}